# A Comprehensive Analysis of Allostery in 14-3-3 ζ Docking Proteins using the Spatial Convolution Model (SCM)

Leroy K. Davis [1]

1-Gene Evolution Project, LLC, lkdavis.geneevolutionproject@yahoo.com, 1- (225) 610-6203

**Abstract:** The Spatial Convolution Model (SCM) analyzes allostery based on the spatial evolution of the docking protein elastic media, whereby convolution of the media in response to wave propagation is solved as a function of Z fluctuations and backbone vibration modes. We show that although the elastic media is a complex three-dimensional structure allostery behaves as if it occurs along a stretched oscillating string, where inhomogeneities along the string effect local entropies responsible for ligand binding and transduction of allosteric waves. To identify inhomogeneities along the string, we ignored local density and tension changes during wave propagation and resolved helix wave and physical properties by applying molecular string and beam bending theories. Importantly, we show that allostery occurs at three major scales and that propagation of standing waves create a rolling entropy which drives entropy transfers between fields. Conversion of resonance energy to quantum harmonic oscillators allowed us to consider effects of damping and interactions with the surrounding media as well as to model effects of residue interaction strength on entropy transfer.

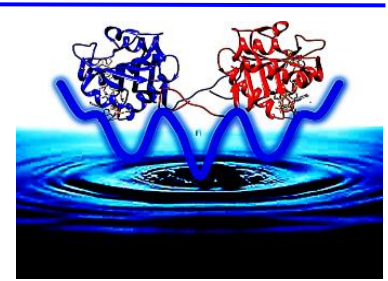

## 1. INTRODUCTION

Allosteric interactions promote cooperative communication between protein domains and the active site by binding of effector molecules at distant sites leading to altered active site geometry and activity. Allostery can effectively govern signal transduction as proteins exist in a population of pre-existing conformational states where each conformer has a different ligand binding potential, thusly allowing for the processing of diverse signals by regulating binding affinities. [1-3] Due to allosteric cooperativity their binding constants vary from apo-proteins to their intermediates, whereby binding of effector molecules make the protein more assessable for downstream binding partners. Early allostery theories were based on two prevailing views conformational selectivity and induced fitting that continue to persist in modern theory. The idea of induced conformational change was introduced by Koshland in his 1958 model that also suggested that proteins exist in conformational ensembles. [4] It was suggested by Kumar in 2000 that opposed to induced conformational change due to sequential binding of ligands on the protein surface, [3] allostery occurs by shifts in the population of conformational ensembles based on the funnel model. [5,6] Theoretical binding funnels are similar to protein folding funnels described in Bryngelson, [5] where an ensemble is a group of closely related conformational states available for binding. While funneling makes the folding process resistant to mutation, Nussinov suggested perturbation at any site can lead to a shift in the distribution of conformational states across the population [1,7]. Prominently, covalent modifications such as mutations may lead to redistribution of the native ensemble of enantiomers, where any perturbation would necessarily involve reorganization of the evolutionarily tuned interaction network. In some cases, this may manifest in diseases such as cancers where the protein is locked into one conformational state and is either constitutively active 'On' or inactive 'Off'. [7, 8]

In recent decades, there have been a broad range of reports on allostery with many addressing roles of entropy and free energy landscapes such as McLeish [9] and Hacisuleyman

[10]. In the current study, we used a spatial convolution model (SCM) to investigate allostery in a synthetic 14-3-3 ζ docking protein engineered in Davis 2019 [11] using the 'Fundamental Theory of the Evolution force' described in Davis 2020 [12]. The technique identifies genomic building blocks or highly fit haplotypes and constructs proteins in a domain Lego fashion by their assembly. This results in the evolutionary conservation of functionally critical gene regions and the insertion of evolution tested sequence blocks taken from orthologue genes. These sequences are inserted into less fit faster evolving regions of the protein allowing us to investigate effects of evolutionally '*safe*' mutation. The SCM permitted us to describe effects of mutation on motion, resonance, and damping; emphasizing that entropy landscapes dynamically change as a result of energy exchanges occurring at different spatial scales.

The SCM establishes helical resonance and damping as critical components of allostery. As Hamiltonians decay due to damping energy is lost at one spatial scale and transferred to a lower scale resulting in a change in entropy and free energy landscapes. We suggest that energy transfer in docking proteins occurs on three major scales characterized by (I) a macroscopic scalar field $E_{Ma}$ generated by helical power strokes during conformational switching that transfers energy to the protein surface and distorts the elastic media, (II) a microscopic scalar field $E_{Mi}$ created by the fast decomposition of the macroscopic scalar field at the end of conformational switching, where the decay of structural oscillations allow energy transfer to the (III) sub-microscopic scalar field whose gradient supplies driving force for thermal fluctuations in the media that characterize motion free allostery. During the process energy is transferred to surrounding fluid as Eddy and Brownian motions as well as sound waves creating an incoherent drive that reaches thermal equilibrium with oscillating helices and affects entropy and free energy landscapes. Thusly, the SCM suggests that opposed to selectivity occurring due to sequential binding, it is purely a function of the changing energy landscape at different scales. These processes have a time-spatial evolution that change the shape and barriers of binding funnels. Thusly, they continuously alter the population of conformational enantiomers, whereby different ligands are optimally preferred at varying time-spatial evolution points. Effector molecules affect binding of ligands by changing the vibrational status and damping rates of helices thusly also altar the entropy landscape.

14-3-3 docking proteins are key regulators that have been reported to interact with over 300 protein binding partners and multiple signaling pathways, including those associated with cell proliferation, cell cycle and apoptosis. [13, 14] The 14-3-3 docking protein family includes multiple isoforms including (β, ε, η, γ, τ, σ, ζ). The focus of the current study is 14-3-3 ζ due to its prominent role in signal transduction and human disease. It has been suggested that targeting of 14-3-3 ζ docking proteins may facilitate a common therapeutic approach against aging, neurodegenerative disease, and cancer. [14-16] The ζ isoform exists both in monomeric and dimeric forms under biological conditions and binds ligands in both states. [17] 14-3-3 ζ also forms tetrameric and octameric biological complexes such as those formed with serotonin N-acetyl transferase. [18] Ligand interactions occur at three binding sites within the amphipathic groove, a highly conserved region characterized by spatially opposed hydrophobic and charged residues. [19] Mutations in this region have been shown to disrupt ligand interaction. [19] We reported the effect of synthetic evolution on this region in Davis 2019 [11] and Davis 2020 [12]. Due to their many binding partners [14] and the presence of multiple binding sites, it is apparent that 14-3-3 ζ consists of a complex interaction of conformational enantiomers. Importantly, the ability to rewire allostery without disrupting native ensembles would allow for altering of cellular pathways and offer the potential for novel drug discovery.

We investigated allostery by modeling docking proteins as being comprised of a continuous homogeneous and isotropic elastic media. This allowed us to isolate the effects of structural inhomogeneities on information transfer while ignoring effects of changing densities and strains during motion. Convolution of the continuous elastic media was modeled as a function of Z fluctuations in respect to backbone vibration modes. Whereby, Z fluctuations estimate the dimensionless moment of residue motions that occur in response to perturbation at the effector site. Le' s plucked molecular string [20] was applied to describe motion in addition to beam bending theory to resolve helix wave and physical dynamics. [21-24] Allosteric radiation describing the flow of information was modeled by mapping Hamiltonians describing helix resonances to quantum harmonic oscillators [25, 26] and incorporating damping effects in respect to interaction with the surrounding fluid. We also incorporated residue adjacency to describe the effects of residue interaction strength on information transfer. By applying SCM, we found that allosteric information was transferred by standing waves and because 14-3-3 ζ is composed of helix-loop-helix domains and lack $\beta$-sheets behaved as if it were being conducted along an oscillating stretched string. Notably, allostery occurred by evolutionally optimized energy transfer through structures and harmonic tuning of resonances that in conjunction with helix orientation promoted the formation of potential wells between domains allowing for trapping of ligands.

## 2. Methods

Normal mode analysis was performed on synthetic docking protein SYN-AI-1 ζ and native *Bos taurus* 14-3-3 ζ monomers and homodimers utilizing the elastic network model (ENM), with a minimum and maximum DQ amplitude perturbation of 100 and a DQ step size of 20. [27, 28] Allosteric motions were approximated by Z fluctuations of residues generated by motion of sites $x_1$ and $x_2$ to a perturbation at an effector residue with an arbitrary minimum distance cutoff of 12 Å. Z fluctuations were solved as the difference in the magnitude and net motion of the elastic media (1). Where, $mag_d$ captures the amplitude of the motion as well as stretching of the media (2). $net_d$ reflects distribution and skewing of the elastic media (3). $L_I$ describes the largest distance increase in response to perturbation at the effector site and $L_D$ the largest decrease. Distance estimations were based on elNemo motion perturbed PDBs that estimate normal modes as eigenvectors of Hessian matrices. ENM was utilized in this study as it estimates motion by superposition of normal modes around a minimum energy conformation, [27, 28] which agreed with SCM which solves for Hamiltonians associated with allosteric motions as functions of fluctuations around minimum energy conformations.

$$Z = mag_d - net_d \qquad [1]$$

$$mag_d = \sum_{r=n}^{N}(|L_I| + |L_D|) \qquad [2]$$

$$net_d = \sum_{r=n}^{N}(|L_I| - |L_D|) \qquad [3]$$

## 3. Results

**3.1 Cooperative Motions in 14-3-3 ζ Homodimers.** We found that in native 14-3-3 ζ homodimers Z fluctuations gave a residual imprint of helix shape and motion vectors **Fig. 1**a,b, helices are denoted by asterisks and coils by arrows. Allosteric wave amplitudes were proportional to those located in adjacent nodes and a function of distance $l$ between wave boundary residues that defines the limit of helix-loop-helix structural domains. Based on Z fluctuations, residues located in the interior of structures were more flexible than those located on exteriors suggesting hinge-bending. Boundary residues flanked domains in three-dimensional space horizontally and vertically, suggesting they are antinodes. Based on the observed boundary conditions and dependence of wave amplitudes on distance $l$, allosteric waves traverse 14-3-3 ζ docking proteins as standing waves that can be described by a Fourier series **(4)**.

$$\omega(x,t) = \sqrt{\frac{2}{l}} \sum_{n=1}^{\infty} sin\frac{n\pi}{l} A(a_n cos\omega_n t + b_n sin\omega_n t)$$

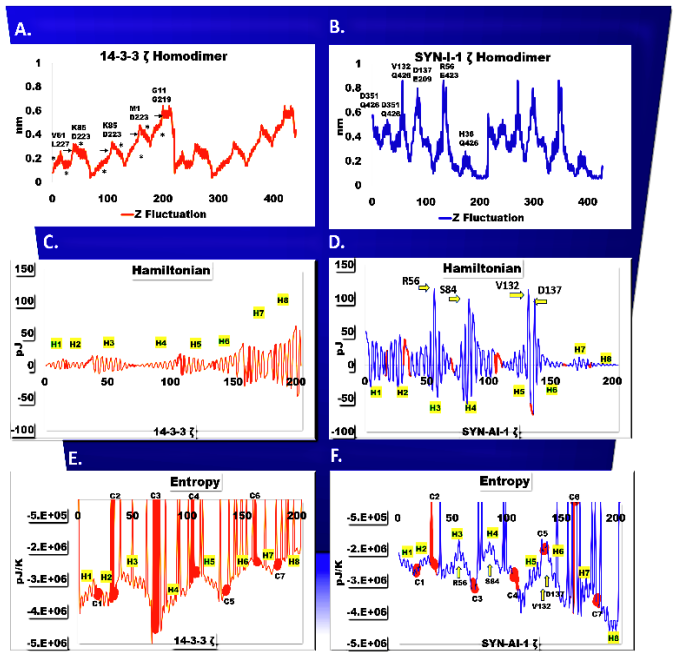

**Fig. 1 Transduction of allosteric Information**

We observed crosstalk between structures as colocalization of helices in wave nodes implies their cooperative motion. This is an important aspect of allostery as the relative motion of domains help shape the active site. [29] Monomers of the native docking protein comprised of two deformation waves an N' terminal wave having two nodes encompassing helices H1 – H3 and a second wave comprised of three nodes spanning helices H4 – H9 **Fig. 1**a. SYN-AI-1 ζ consists of a single deformation wave with a superposition of six smaller waves, each associated with an individual α-helix suggesting they work cooperatively in one motion **Fig. 1**b.

**3.2 Sequence Dependent Effects.** Interactions between effector sites located at Z fluctuation amplitudes and allosteric sites that respond to their perturbation are shown in **Fig. 1**a,b. We found that in the SYN-AI-1 ζ homodimer, allosteric pathways were routed through helices H3 and H4 as the motion of effector residue R56 located in the H3 interior was allosterically coupled to V132 located at the H5 C' terminal as well as to residue Q426 located in helix H16 of the adjacent monomer. Additionally, residue S84 located in helix H4 was allosterically coupled to residues D137 of helix H5 and E209 of helix H8. There was strong sequence dependence in the native 14-3-3 ζ homodimer as 4 of 5 wave amplitudes were associated with amphiphilic residues (Q, N, K) and 5 of 10 allosteric sites with amphiphilic residues (K, D). Sequence dependence was also strong in SYN-AI-1 ζ as 3 of 6 wave amplitudes were associated with amphiphilic residues (E, N, R). The remaining sites were occupied by residues (M, V, S) that have sidechains of low volume that increase rotational freedom.

**3.3 Scales of Allosteric Energy Transfers**
**3.31 Allostery at the Macroscopic Scale.** ENM simulations suggests that motion of structural domains occurred in spatially and temporally discrete steps that limit decoherence. Thusly, allostery was modeled as an energy transfer process across a continuous multidimensional elastic media that is isotropic and homogeneous having a uniform tension $\tau = 1$ and density $\mu = 1$, where mass $m = \mu \Delta l = 1$. SCM describes allostery based on spatial evolution of the continuous elastic media in response to perturbation. Motion is solved as a function of Z fluctuations in respect to tangential $\Delta z$, rotational $\Delta \theta$, and stretching $\Delta l$ modes, where $\Delta \vec{r}$ is the resultant of backbone vibration modes. Thusly, we account for convolution of the elastic media located between effector and allosteric sites.

$$K = \frac{1}{2} \mu \Delta l \left( \frac{\partial Z}{\partial \Delta z, \partial \Delta \theta, \partial \Delta l} \right)^2 \Delta \vec{r}^2 \qquad [5]$$

$$U = \frac{1}{2} \tau \Delta l \left( \frac{\partial Z}{\partial \Delta z, \partial \Delta \theta, \partial \Delta l} \right)^2 \Delta \vec{r}^2 \qquad [6]$$

As standing waves traverse the elastic media, they displace residues creating local volume and density changes as well as creating local energy, pressure, and temperature gradients. [30] Due to inhomogeneities between helices as well as between helices and coils, these effects vary across the docking protein and affect the entropy landscape. By ignoring effects of density and tension, we were able to eliminate local variances in residue mass and C-alpha strain allowing us to isolate the effects of inhomogeneities in the continuous media on allostery (7).

$$H \Rightarrow \sum_{res=1}^{helix} l \left( \frac{\partial Z}{\partial \Delta z, \partial \Delta \theta, \partial \Delta l} \right)^2 \Delta \vec{r}^2 \qquad [7]$$

Based on wave Hamiltonians, allostery is a function of damped harmonic oscillations **Fig. 1**c,d. Spikes in the Hamiltonian near effector residues R56 and S84 of SYN-AI-1 ζ corroborate that allosteric pathways were routed through helices H3 and H4. We also observed spikes near allosteric residues V132 and D137 supporting that amphiphilic residues and residues of low volume promote allostery. Entropy $S = -0.5 K_B T Log H$ within the elastic media was solved as a function of standing wave Hamiltonians **Fig. 1**e,f. We found that docking proteins were structured so that orientations of helix-loop-helix domains and the direction of harmonic decay created entropy wells that promote allostery and the flow of entropic information. In the native homodimer, entropy wells were created by combined motions of helices H1 and H2, H2 and H3, H4 and H5, and H6 and H7. There was some disruption of this trend in synthetic homodimers, where entropy wells did not form exclusively by combined Hamiltonians of helices but formed independently in helices H3 and H4 due to insertion of amphiphilic sequence blocks. Both proteins formed entropy gradients from their N' to C' terminals. However, direction of the gradient was reversed in SYN-AI-1 ζ. [26] There was a strong association of coils and entropy as interdomain coils C2, C4, and C6 were all located in entropy wells and exhibited steep entropy increases $S \rightarrow 0$, whereby intradomain coils C1, C5, and C7 were entropically stable **Fig. 1**e. This held true with some deviation in the SYN-AI-1 ζ homodimer, as interdomain coils C2 and C6 were associated with entropy wells and steep entropy increases $S \rightarrow 0$, while intradomain coils C1, C3, C5 and C7 were entropically stable **Fig. 1**f. Our findings support that entropy is a function of mobility, [10] where allostery was routed through interdomain coils to allow rotational freedom during conformational switching. [29]

The large energy gradients in Hamiltonians characterizing helix-loop-helix structural domains imply that conformational switching associated with helix power strokes occurs in discrete energetic steps as suggested by Huang. [31] Driving force of the macroscopic step is given by its Langevin (8). Where, partial derivatives of Z fluctuations were solved in respect to backbone vibration modes, $\Gamma$ is the damping coefficient and the last term gives harmonic potential.

$$F(Z) = m[\partial^2 Z + \Gamma \partial Z]_{\Delta z, \Delta \theta, \Delta l} + \nabla_Z H_{Ma} \qquad [8]$$

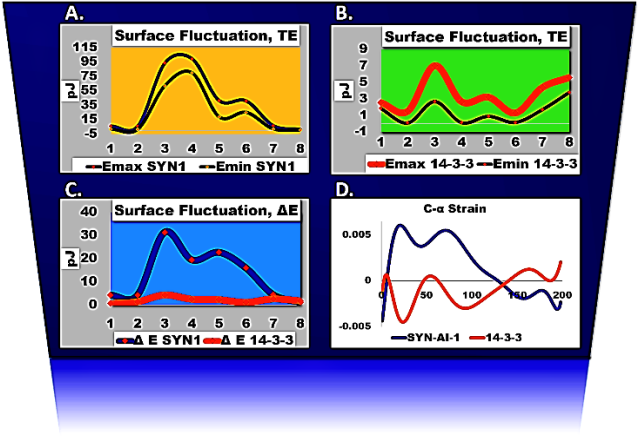

**Fig. 2 Surface Energy vs. C-alpha Strain**.

Fast decomposition of the macroscopic scalar field $H_{Ma} = \Psi_H + (\phi + \Omega)$ creates internal energy $\Psi_H$ and flux through fast energy channels to the microscopic scalar field $|\phi + \Omega|H_{Mi}$. Hamiltonians $(\phi + \Omega)$ describe energy transferred to helix oscillations $\phi$ and surrounding fluid $\Omega$, whose gradient $\nabla_{\phi, \Omega}$ describes driving force through lower energy channels. Due to efficiency of coherent energy transfer by waves, [32] the Hamiltonian of the macroscopic field $H_{Ma} \rightarrow \Delta E = E_{max} - E_{min}$ is described by $\Delta E$ the difference in helix maximum and minimum energy conformations which describes the energy produced during the conformational switch. Damping affects the Hamiltonian by $H_{Ma} = \Delta E - \Delta E_\Gamma$, where energy $\Delta E_\Gamma$ loss due to damping $\Delta E_\Gamma = (m\omega_0^2 Z_0^2/2) \cdot e^{-\Gamma^2}(1 - e^{-\Gamma \tau_0})$, and $\tau_0$ is the oscillation period. Notably, a correlation exists between $\Delta E$ and C-alpha strain suggesting that allostery depends on the potential energy landscape **Fig. 2**. The relationship is not exact as $\Delta E$ is the summation of residue energies in helices and does not incorporate motion of coils, whereby C-alpha strain was shown continuously across each residue.

According to the Fluctuation-Dissipation Theory, energy gradient $\nabla_z H_{Ma}$ of the macroscopic field may be divided into a driving force $F(\Psi_H)$ which is dependent on the energy gradient and a dissipation force $f(\phi + \Omega)$.

$$\nabla_z H_{Ma} = F(\Psi_H) + f(\phi + \Omega) \qquad [9]$$

Internal energy $\Psi_H = H_{Wave} - E_\Omega$ is the energy absorbed from the standing wave into helices described by standing wave Hamiltonian $H_{Wave} = l(\partial Z/\partial \vec{r})^2 \Delta \vec{r}^2$ **Fig. 1**c,d and energy $E_\Omega$ donated to surrounding fluid during confirmational switching. $\Psi_H$ is a function of the state of the oscillator as well as its internal heat and work which are connected to its state by the First Law of Thermodynamics $d\Psi_H = dQ + dW$. [30] Work $W = \Delta \vec{r} F(Z)$ is a function of driving force and the resultant of instantaneous backbone vibration modes. During standing wave propagation, due to speed of the conformational switch this relationship is shifted toward work. At completion of the helix power stroke internal energy is converted $|\Psi_H|\phi + \Omega$ to the microscopic field and equilibrium $Q \leftrightarrow \phi$ reached between structural resonances and thermal fluctuations in the elastic media and surrounding fluid. Heat generated by oscillating structures can be described by resonance free energy $G_\phi = k_B T \ln\{1 - \exp[-\phi /(k_B T)]\}$. [33] Energy $\Omega$ is transferred to fluid as shock waves caused by the sudden stopping motion at the end of the conformational switch and equilibrium reached $\{\Psi_H, \phi\} \leftrightarrow \Omega$ between the media and oscillating helix. Energy is exchanged with surrounding fluid as a function of helix driving force and damping in conjunction with the incoherent driving force of the surrounding fluid.

At completion of the conformational switch driving force is no longer a function of the Langevin given in (8), but due to decomposition of the system $F(\Psi_H)$ becomes a function of the harmonic gradient $\nabla_z H_{Ma}$ characterizing helices as defined by the third term of equation (8), and the Onsager relationship, [34, 35] defined by $\langle \Delta Z^i \Delta Z^j \rangle = K_B T(\partial M_Z/\partial F^j)$ which describes harmonic fluctuation about the minimum energy conformation. Where, $F^j$ the force occurring at helix end $j$ is a function of N' and C' terminal instantaneous Z fluctuations $(\Delta Z^i, \Delta Z^j)$ about time-averaged positions of residues described by dimensionless second moment, $M_Z$. By rearrangement and integration $F(\Psi_H)$ is given by (10), where element $i$ gives instantaneous force at any given position within the helix.

$$F_i(\Psi_H) = -\frac{1}{N} K_B T \sum_{i=1}^{N} \frac{M_Z^i}{\Delta Z_{flux}^i} \qquad [10]$$

Synthetic evolution generated changes in driving force $F_i(\Psi_H)$ due to introduction of disordered regions near effector residue R56 of helix H3 and effector site S84 of helix H4 **Fig. 3**a. The loss in driving force in these regions correlated with entropy wells shown in **Fig. 1**f, that promoted the routing of allosteric pathways through helices H3 and H4. These regions displayed low mechanical stiffness supporting that areas of low stiffness are involved in transfer of allosteric information **Fig. 3**c. Symmetry breaking in these regions was supported by higher mechanical strains $\varepsilon$ due to longer backbone stretching modes **Fig. 3**e. Notably, harmonic decay of driving force from N' to C' terminals in native helices **Fig. 3**b, signified genomic optimization of allostery by oscillatory damping. SYN-AI-1$\zeta$ displayed some loss of tuning in helices H3 and H4 suggesting that native 14-3-3 $\zeta$ docking proteins are evolutionarily tuned **Fig. 3**a. However, the loss in driving force in helices H3 and H4 produced minimum changes in harmonic damping **Fig. 4**a suggesting that the system is very robust. Importantly, decay of driving force $F(\Psi_H)$ toward helix-to-coil transitions supports that coils play prominent roles in allostery.

$$C_{\vec{r}}(Z) = \langle \Delta \vec{r}(Z') \Delta \vec{r}(Z' + Z) \rangle \qquad [11]$$

$$\tilde{C}_r(Z) = K_B T \frac{\Gamma^2}{H(1 + \Gamma^2)} \qquad [12]$$

Correlation function $C_{\vec{r}}(Z)$ was solved according to Pleiss [36] but in the spatial dimension (11). Spectral oscillations are described in (12). Autocorrelation revealed increased damping in helices $H5 - H8$ of SYN-AI-1 $\zeta$ that indicate greater energy dissipation and a faster decay rate into thermal fluctuations. Changes in the $\Gamma$ direction indicate redistribution of thermal fluctuations that effect local entropies in surrounding media which affect binding affinities of downstream ligands. Allosteric rate $A = \Gamma^{-1}/\Delta \vec{r}$ indicated a significant loss of information at C' terminals in the synthetic homodimer due to disruption of harmonic tuning and increased stiffness **Fig. 4**b. Additionally, Fourier Transform $\tilde{C}_Z$ indicated inversion of the docking protein motion vector **Fig. 4**c. It is worth mentioning that in respect to $\nabla_Z H_{Ma}$, the motion equation becomes $\ddot{Z} + \Gamma \dot{Z} + \omega^2 Z = 0$ a differential equation that is a function of the dimensionless moment of the elastic media and velocity of the plucked molecular string, where $\omega$ captures convolution of the continuous elastic media.

### 3.32 Allostery at the Microscopic Scale

Microscopic scalar field $H_{Mi}$ has a separate Langevin (13), whose gradient $\nabla_Z H_{Mi}$ describes driving force $F(\phi, \Omega)$ of structural oscillations as well as Brownian and Eddy motions in the surrounding fluid. Spatial decay $\zeta_{\phi,\Omega} = \nabla_Z \times H_{Mi}$ describes energy flux to lower channels that produce thermal fluctuations in surrounding elastic media that result in a motion free allostery that does not rely on conformational switching but exploits thermal consequences of entropy $S = -0.5 K_B lnk$. [9]

$$|\dot{\phi} + \Omega|H_{Mi}\rangle \Rightarrow \nabla_Z H_{Mi} = F(\phi, \Omega) + \zeta_{\phi,\Omega} \qquad [13]$$

Energy $\Omega$ donated to surrounding fluid is in constant flux and is the difference in the macroscopic field Hamiltonian and helix internal energy and resonance which undergo continual decay $\backsim_{\Psi,\phi}$ into surrounding media.

$$\Omega = H_{Ma} - (\Psi_H \leftrightarrow \phi) + \backsim_{\Psi,\phi} \qquad [14]$$

In addition to decay of the microscopic field, shock waves following helix power strokes interfere with longitudinal waves generated during conformational switching thusly resulting in decoherence and formation of vortical motions that contribute to driving forces responsible for thermal fluctuations. Although, resulting entropy changes due to fluid $\Omega$ effects altar ligand binding affinity, when modeling allostery we focused on the effects of helical resonance $\phi = l\omega^2 \Delta \vec{r}^2$. Whereby, standing wave angular velocity $\omega$ was solved according to Le's plucked molecular string solution (15). We modeled the system from the perspective that as waves propagate the elastic media each residue acts as a deflection point characterized by an initial velocity of $\omega_0 = M_Z \sin(\Delta\theta) \cos(\Delta\theta)$ that is a function of its time-averaged position $M_Z$ and instantaneous rotational change $\Delta\theta$ about peptide bonds. Final velocities $\omega(\Delta \vec{r}, t)$ occur at wave amplitudes and are a function of Z fluctuations. Thusly, we were able to model convolution of the elastic media during wave propagation using Le's solution.

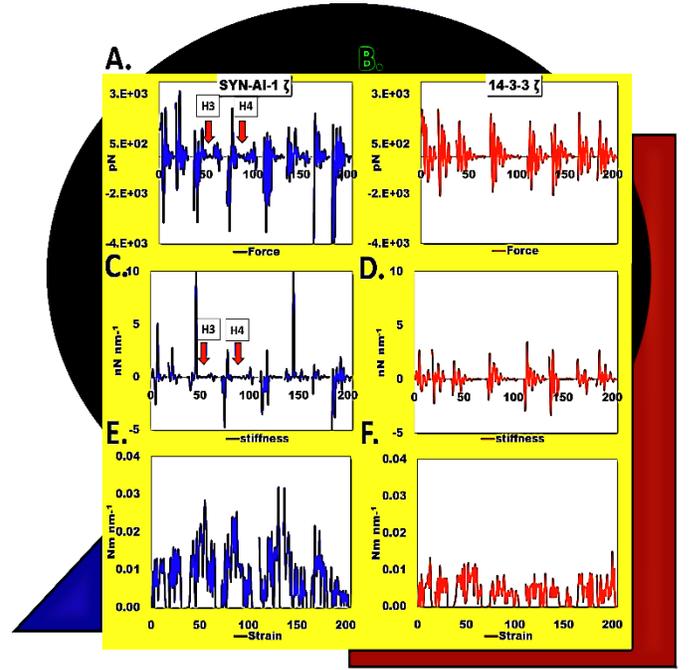

**Fig. 3 Helical Dynamics.**

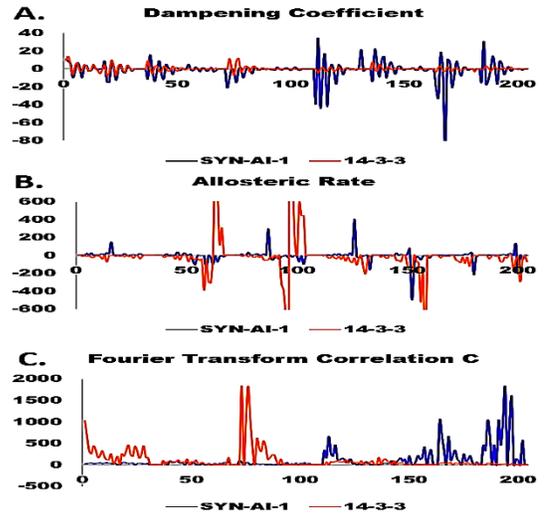

**Fig. 4 Damping Effects**

$$\omega(x,t) = \frac{2l^2 \omega_0 \sin\left(\frac{\pi a n}{l}\right)}{\pi^2 n^2 a(1-a)} \sin \frac{n\pi x}{l} \cos \frac{n\pi c}{l} t \qquad [15]$$

Le's solution was very sensitive to deflection points thusly precise deflection points were solved by equilibrating $\phi$ with wave Hamiltonians solved by spatial evolution of phase space $\partial M_Z / \partial \Delta \vec{r} \Rightarrow \omega(Z)$ (16). We confirmed the two were equivalent with the exception that the initial $\phi$ was out of phase because we estimated deflection points. The Fourier was simplified by solving for motion orthogonal to the x-axis at $w_0 \sin(\pi/2) = 1$ and incorporating harmonic deflection point $\vartheta = (2\pi \bar{r}_n/l)(1 - 2\pi \bar{r}_n/l)$, where $\bar{r}_n$ is the residue number. We made the Fourier time-independent which allowed the Hamiltonian to be solved solely by spatial evolution of the resonance field. By integrating $\omega(Z)$ (17), Z fluctuations were approximated at an $R^2=0.9987$ and a variance of 5.24105E-06 **Fig. 5**a.

$$\frac{\partial M_Z}{\partial \Delta \vec{r}} \Rightarrow \omega(Z) = \frac{2l^2}{\pi^2 n^2 \vartheta} \sin\frac{n\pi\Delta\vec{r}}{l} \cos\frac{n\pi\Delta\vec{r}}{l} \qquad [16]$$

$$Z_{flux} \Rightarrow \int \omega(Z) d\Delta\vec{r} = \frac{2l^4}{\pi^4 n^4 \vartheta} \left( \cos\frac{n\pi\Delta\vec{r}}{l} \sin\frac{n\pi\Delta\vec{r}}{l} \right) \qquad [17]$$

By modeling helical resonance ϕ as described, we were able to observe linear effects of residue deflection on standing wave velocity **Fig. 5**b, in addition to harmonic fluctuations about minimum energy conformations **Fig. 5**c,d. Deflection slopes captured acceleration effects due to motion mechanisms and indicated hinge-bending motions of helix H3 and H4 are slower than neighboring gating motions. The magnitude of harmonic fluctuations correlated with standing wave dynamics reported in **Fig. 1**a,b, and also captured spikes in motion. However, differences in energy transfer characterized by increased wave velocities through helices H3 and H4 could not be explained by wave dynamics **Fig. 5**e,f, but correlated directly with regions of symmetry breaking.

Additionally, we found that the pattern of resonance energy ϕ resembled the macroscopic field Hamiltonian $\Delta E$ **Fig. 6**f,g. This is theoretically correct as $\Delta E$ is a superimposition of ϕ due to energy transfer at conclusion of the conformational switch. Notably, ϕ also resembled C-alpha strain **Fig. 6**a,d. These data combined with behavior of standing wave Hamiltonians and driving force suggests that although the elastic media of docking proteins is a complex multi-dimensional structure allostery behaves as if it occurs along an oscillating stretched string. Based on docking protein entropy landscapes and wave velocities through helices **Fig. 1**e,f and **Fig. 5**e,f, the transfer of allosteric information depends on inhomogeneities within the string. Importantly, this was supported by the fact that standing wave velocities through helices comprising helix-loop-helix domains were of a similar magnitude. The helices have similar dynamics and resonance patterns due to their coevolution which allows balanced and optimal energy transfer through structural domains. This was true at the macroscopic scale as depicted by resonance patterns of helix Hamiltonians **Fig. 1**c,d and also held true for resonance ϕ at the microscopic scale, particularly in (H1, H2) and (H7, H8) helix groupings **Fig. 5**c,d and **Fig. 6**a,b. While the magnitude of ϕ was similar for helices H3 and H4, the lower energy of H4 was only due to its dorsal location to helix H3 which restricted its hinge-bending motion. This was also true of helices comprising the H5-C5-H6 domain.

### 3.4 Effects of Mechanical Dynamics on Allosteric Motion and the Entropy Landscape.

To investigate helix mechanical dynamics that contribute to inhomogeneities in the oscillating stretched string, we solved elastic moduli by assuming potential energy $U_{\Delta\vec{r}}$ associated with convolution of the elastic media during allostery was equal to elastic potential energy $U_{elastic}$

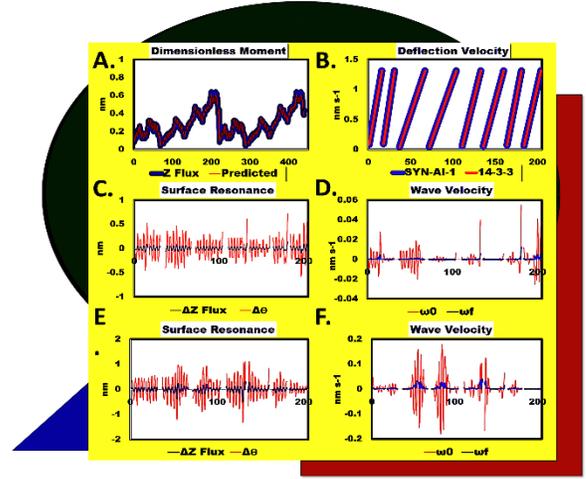

**Fig. 5 Motion of the Elastic Media in Response to Perturbation.** Z fluctuation predicted vs. experimental (**A**), Deflection velocity of helices (**B**). Resonance ϕ energy 14-3-3 ζ (**C**) and SYN-AI-1 ζ (**D**). Wave velocity through helices of 14-3-3 ζ (**E**) and SYN-AI-1 ζ (**F**).

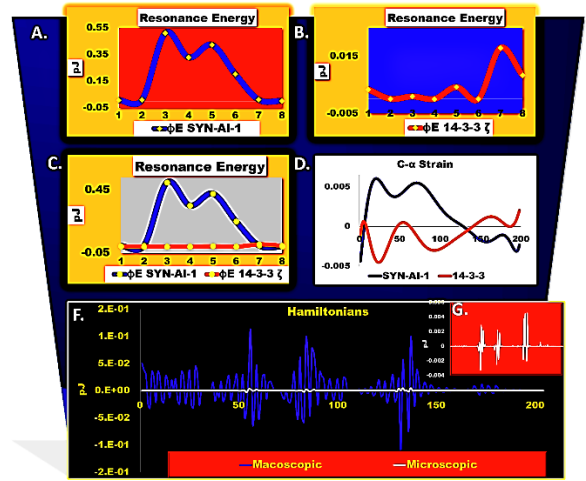

**Fig. 6 Resonance Energy vs. C-alpha strain.**

associated with mechanical stretching of structures according to the nonlocal theory of elasticity. [24, 37]

$$\overbrace{\frac{1}{2}l\left(\frac{\partial Z(\Delta\vec{r},t)}{\partial \Delta\vec{r}}\right)^2 \Delta\vec{r}^2}^{U_{\Delta\vec{r}}} = \overbrace{\frac{1}{2}\frac{\lambda A_0}{l}\left(\int_0^L \frac{\partial A_b(\Delta\vec{r},t)}{\partial \Delta\vec{r}}\right)^2}^{U_{elastic}} \qquad [18]$$

Hookean spring constants were solved as a function of the response of the continuous elastic media to propagating waves, whereby resolving equation (18) we described elongation $\Delta l = 0.9129\omega\vec{r}\sqrt{l/\lambda}$ of structures as a function of wave velocity and the structure's mechanical resistance $\lambda$. Allosteric stretching increases with the length to stiffness ratio and vibration mode. Thusly, SCM characterizes allostery by describing interaction of waves with the media. The effects of allosteric stretching on the elastic media were captured by elongation spring constant $k_l$ (19), which we solved by modeling the elastic media as an oscillating (LHS), stretchable (RHS) molecular string with a resonance amplitude of $A = \int \omega(\Delta z)$ and a stretching motion of amplitude $A_{\Delta l} \equiv \Delta l$.

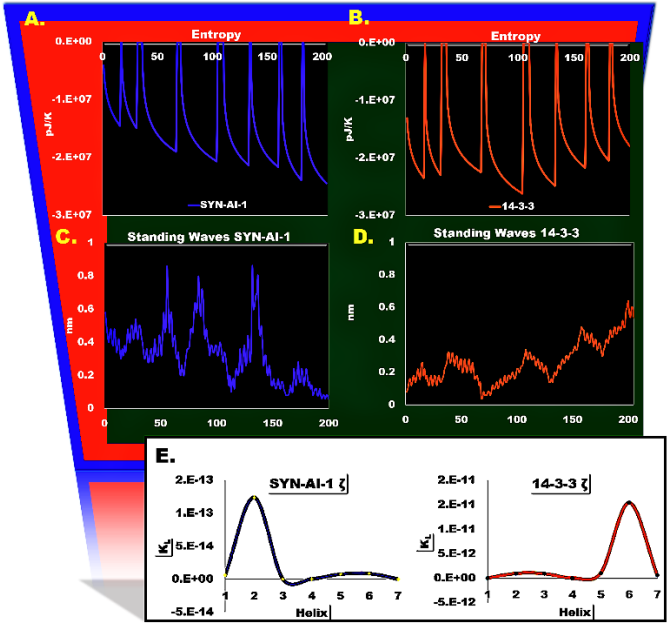

**Fig. 7 Standing Wave Effects on Entropy and Plasticity**

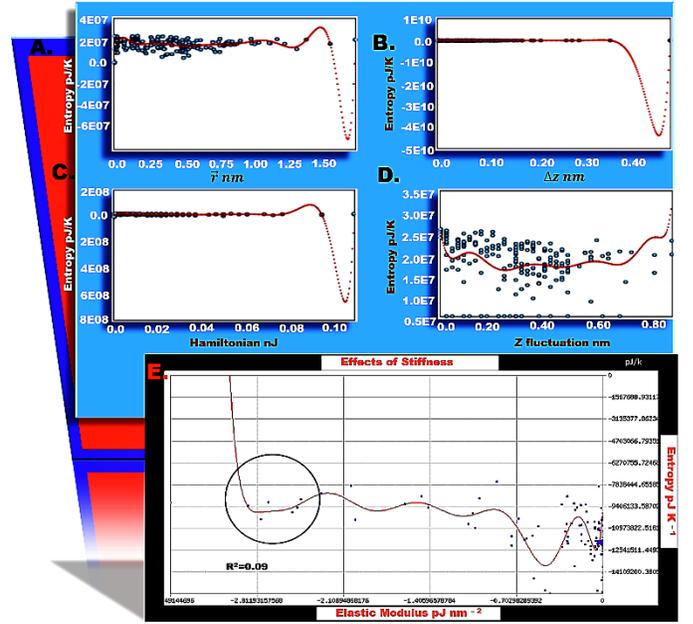

**Fig. 8. The Standing wave Propagation vs. Rolling Entropy Landscape.** The Blue.Stats online polynomial regression tool was utilized to compare the rolling entropy landscape to residue backbone stretching mode $\vec{r}$ (**A**), instantaneous $\Delta z$ fluctuations (**B**), the standing wave Hamiltonian (**C**), surface Z fluctuations (**D**), and elastic modulus k (**E**). All models were confirmed with p-values=0.

$$\frac{2l^4}{\pi^4 n^4 \vartheta}\left(cos\frac{n\pi\Delta r}{l}sin\frac{n\pi\Delta r}{l}\right) = \overbrace{0.9129\,\omega\vec{r}\,\sqrt{l/\lambda}}^{A_{\Delta l}} \qquad [19]$$

Allostery at the macroscopic scale was characterized by a rolling entropy landscape that dynamically changed as waves propagated across the protein. Elastic modulus $k_l(motion) \Rightarrow [\pi^4 n^4/l^3]_{\Delta \vec{r}^2}$ captures dynamic changes in the elastic media in response to wave propagation as a function of residue backbone vibration modes. Elastic modulus of motion $k_l(motion)$ was derived by substituting the Fourier solution for velocity (16) into equation (19). Entropy $S = 0.5K_B TLNk_l(\text{motion})$ gave a snapshot of the rolling entropy landscape traveling the surface with standing waves **Fig. 7**a,b. Based on the location of entropy nodes in respect to Z fluctuations **Fig. 7**c,d, valleys located between standing wave nodes characterize free energy $\Delta G$ wells that formed across the protein surface by combined motions of helices about coils during conformational switching **Fig. 7**c,d. Barrier height and well depth were functions of Z fluctuations of interacting helices, and well width a function of the distance between cross talking helices.

$k_l(motion)$ depends on the flow of information and was not significantly affected by damping. Limited effects of damping were reflected as curves located at apexes of entropy nodes that occur as inverse functions of the exponential decay of helix resonances and reflect the interface of microscopic and macroscopic fields. Based on the orientation of entropy nodes to Z fluctuations, they were formed by barriers of $\Delta G$ wells. Notably, by observing the entropy landscape we noticed that increased entropies were associated with elevated Z fluctuation amplitudes signifying greater residue motions in response to perturbation. The relationship between media convolution and entropy during wave propagation was confirmed by comparing backbone vibration mode $\vec{r}$ to rolling entropy and resulted in a correlation of $R^2 = 0.55$ **Fig. 8**a, and by comparing rolling entropy to instantaneous Z fluctuations $\Delta z$ at a $R^2 = 0.47$ **Fig. 8**b. We corroborated the relationship by comparing the SYN-AI-1 $\zeta$ standing wave Hamiltonian to rolling entropy at an $R^2 =$ 0.25 **Fig. 8**c. However, we could not directly link Z fluctuations to dynamic entropy likely because entropy nodes form in their valleys and are a function of long-range interactions between helices **Fig. 8**d. Entropy transfer decreased with stiffness of the elastic modulus and was restricted to entropies located in the top quantile **Fig. 8**e (circled). Elastic modulus $k_l = \phi_l/\Delta l^2$ describes stiffness at an instantaneous position $\vec{r}$ during media convolution. Due to its dependence on elongation Hamiltonian $\phi_l$, $k_l$ suggests that while standing waves affect the direction of plasticity, the two-order difference in magnitude results from inhomogeneities introduced by mutation **Fig. 7**e,f.

Elastic modulus $k_\theta = \phi_\theta/\Delta\theta^2$ is a function of rotational Hamiltonian $\phi_\theta$ and describes distortions in the continuous elastic media due to residue twisting modes. $\Delta\theta$ describes the instantaneous change in angles between residues as standing waves traverse helices creating multiple deflection points. Due to nonuniform helical curvatures resulting from varying effects of waves on rotation modes, $\theta$ was solved as a conditional probability based on the motion of the adjacent residue (20).

$$\theta = P\left(\theta_N\mid\theta_{N-1} = \sum_{i=1}^{N-1}\theta_i,\ \ \theta_i = \arctan\left(\frac{\Delta z}{l}\right)\right) \qquad [20]$$

$$\phi = \phi_z + \phi_l + \phi_\theta, \quad \begin{bmatrix} \phi_Z = \frac{k_z - \Delta k_z}{k_0}\phi_\Gamma, & k_Z = \frac{\phi}{\Delta z^2} \\ \phi_l = \frac{k_l - \Delta k_l}{k_0}\phi_\Gamma, & k_l = \frac{\phi}{\Delta l^2} \\ \phi_\theta = \frac{k_\theta - \Delta k_\theta}{k_0}\phi_\Gamma, & k_\theta = \frac{\phi}{\Delta\theta^2} \end{bmatrix} \qquad [21]$$

Hamiltonians $(\phi_Z, \phi_l, \phi_\theta)$ were solved as functions of elastic moduli $(k_Z,\ k_l, k_\theta)$ characterizing the microscopic field and background resonance $\phi$ (21). They were normalized by initial spring constant $k_0 = k_Z + k_l + k_\theta$ and accurate spring constants and Hamiltonians solved by recursion. Elastic moduli associated with helical resonances occurring at the microscopic scale are affected by damping, thusly we considered their time evolution where $\Delta k_Z$, $\Delta k_l$, and $\Delta k_\theta$ are incremental changes due to damping and $\phi_\Gamma$ is the damped resonance energy. These effects were not applied to the initial spring constant $k_0$.

Notably, we show that inhomogeneities introduced into the continuous elastic media by synthetic evolution and captured by elastic modulus $k$ resulted in entropy increases **Fig. 9**c, that correlate with loss of driving force $F_i(\Psi_H)$ within helices H3 and H4 **Fig. 3**a. Entropy wells were associated with areas displaying coil-like behavior located near effector residue R56 in helix H3, and effector residue S84 of helix H4 **Fig. 9**c. Effector sites were associated with decreases in elastic modulus $k$ **Fig. 9**a,b, and were in regions that correlate to increased wave velocities through helices due to symmetry breaking **Fig. 5**f. Synthetic evolution significantly increased plasticity in these regions **Fig. 9**e,f. While entropy changes were more significant in SYN-AI-1 $\zeta$ **Fig. 9**c, entropy wells formed around effector residues N38, R60 and K85 of native 14-3-3 $\zeta$ responsible for naturally occurring allosteric routes in helices H3 and H4 **Fig. 9**d. The latter did not disrupt driving force or helix tuning as they appear to be evolutionarily optimized.

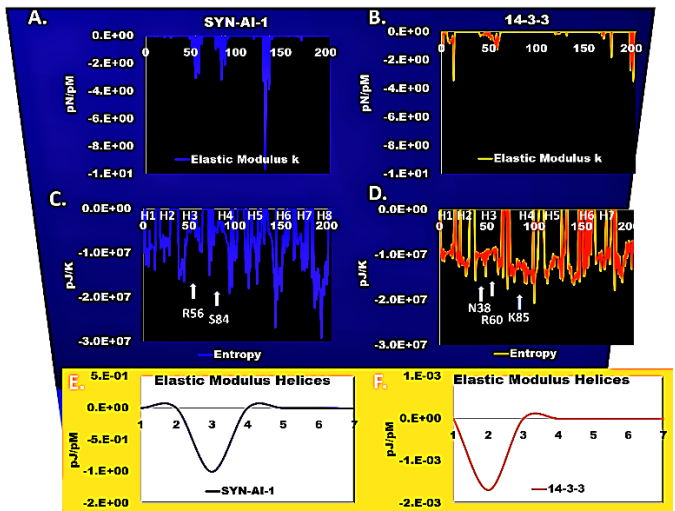

**Fig. 9 Effect of Inhomogeneities in Elastic Media on Entropy.** Etropy $S = -0.5K_B \ln k$ was solved as a function of elastic modulus $k$ of synthetic docking protein SYN-AI-1 $\zeta$ (**A**) and native 14-3-3 $\zeta$ (**B**). Entropy landscapes of SYN-AI-1 $\zeta$ (**C**) and 14-3-3 $\zeta$ (**D**). Elastic moduli $k$ were summed over synthetic (**E**) and native helices (**F**).

### 3.5 Modeling Resonances as Quantum Harmonic Oscillators (QHOs)

It has been shown that energy transfer within continuous collision free media occur on multiple scales from macroscopic to sub-electron. [38] To describe flow of allosteric energy at these scales, structural resonance $\phi$ was converted to QHOs using the time-independent Schrodinger equation (22).

$$\phi \, |QHO\rangle \Rightarrow \varphi = \left(-\frac{\hbar}{2m}\nabla^2 + U(\Delta\vec{r})\right)\phi \qquad [22]$$

$$\nabla^2 Z = \frac{\partial^2 Z}{\partial \Delta z^2, \partial \Delta \theta^2, \partial \Delta l^2} \qquad [23]$$

We approximated $\varphi$ at discrete vibrational states according to [25, 26], where the solution for eigenvalues is given in (24). Notably, we have shown in the previous section that energy transfer $\hbar\omega$ within the elastic media is due to inhomogeneities in the oscillating stretched string that effect wave velocity.

$$\bigwedge = 2n_z + 1 \Rightarrow \varphi = \left(n + \frac{1}{2}\right)\hbar\omega \qquad [24]$$

Following energy transfer at the end of the conformational switch, Hamiltonian $\varphi$ undergoes energy dissipation caused by damping $\Gamma$ of structural oscillations. Thermal equilibrium is achieved by energy transfer between oscillating structures and a surrounding bath having an incoherent drive that is a function of its occupation state and quantum mechanical fluctuations. [39] We modeled energy transfer according to Hauer, [39] and expressed Hamiltonian $\varphi$ using creation $\hat{b}^\dagger$ and annihilation $\hat{b}$ operators that act on eigenstates by $\hat{b}^\dagger |n\rangle = n + 1$ and $\hat{b}|n\rangle = n - 1$ allowing for transformation $|\phi|\hat{b}^\dagger, \hat{b}|\varphi\rangle \leftrightarrow |\varphi|\hat{b}^\dagger, \hat{b}|\phi\rangle$ between resonance $\phi$ and quantum scalar fields $\varphi$ in both directions (25). In the LHS expression, the classic Hamiltonian depends on backbone displacement $\Delta\vec{r}^2$ which is a function of $\hat{b}$ and $\hat{b}^\dagger$. In the RHS, the quantum Hamiltonian is expressed by eigenstate operator $\hat{N} = \hat{b}^\dagger \hat{b} |n\rangle$ that maps these vectors to the quanta number where $\hat{b}^\dagger \hat{b} |n\rangle = (\hat{b}^\dagger \hat{b} + 1/2)$. Our model holds in the classic case as previously described and in the quantum case as we have demonstrated that energy transduction through a homogeneous, isotropic media depends on wave velocity.

$$\frac{p^2}{m} + \frac{1}{2}m\omega_0^2\Delta\vec{r}^2 = \langle\phi|\hat{b}^\dagger, \hat{b}|\varphi\rangle = \hbar\omega_0\left(\hat{b}^\dagger\hat{b} + \frac{1}{2}\right) \qquad [25]$$

where, $\Delta\vec{r}^2 = \left(\sqrt{\frac{m\omega_0}{2\hbar}}\left(\hat{b} + \hat{b}^\dagger\right)\right)^2$

### 3.51 Effects of Surrounding Bath on Hamiltonians

Effects of the surrounding bath are captured by the motion equation $\dot{\hat{b}}_\gamma = -i\omega_0\hat{b}_\gamma - \frac{\Gamma}{2}\hat{b}_\gamma + \sqrt{\Gamma}\hat{b}_n$ introduced in [39], $\hat{b}_\gamma$ is the damped oscillator, the first term on the RHS describes the initial energy of the oscillator, the second term describes energy decay into the bath due to damping $\Gamma$, and the third term describes effects of drive from white noise originating from the bath. [39] When damping coefficient $\Gamma = 0$ structural vibrations are decoupled from the bath. The SCM is helpful in modeling bath interactions as due to interference from white noise a time representation of $\hat{b}_\gamma(t)$ is difficult to obtain over the small bandwidth $\sim\Gamma$ about helix resonance. [39] Therefore, spectral forms of annihilation and creation operators were obtained and solved in respect to velocity of waves through the media, $\hat{b}_\gamma(\omega) = \mathcal{F}[\hat{b}_\gamma(\Delta\vec{r})]_\omega$, $b_\gamma^\dagger(\omega) = \mathcal{F}[b_\gamma^\dagger(\Delta\vec{r})]_\omega$ (26, 27). Decay energy was solved using the classic Hamiltonian as in the case of an oscillating structure the LHS of equation (25) reduces to the second term, [39] thusly the phase shift is $\alpha = \varphi - m\omega_0^2\Delta\vec{r}_\gamma^2$. Amplitude of the damped motion $\Delta\vec{r}_\gamma(\omega) = \sqrt{m\omega_0/2\hbar}\left(\hat{b}_\gamma(\omega) + b_\gamma^\dagger(\omega)\right)$ is the resultant of backbone vibration modes. We ignored mass thusly $\alpha$ is a function of wave velocities through the convoluted media and describes $\hbar\omega$ of decay therefore can be quantized.

$$\hat{b}_\gamma(\omega) = \frac{\sqrt{\Gamma}\hat{b}_n(\omega)}{i(\omega_0 - \omega) + \Gamma/2} \qquad [26]$$

$$\hat{b}_\gamma^\dagger(\omega) = \frac{\sqrt{\Gamma}\,\hat{b}_n^\dagger(\omega)}{i(\omega_0 - \omega) + \Gamma/2} \qquad [27]$$

### 3.52 Effects of Residue Adjacency on Allostery

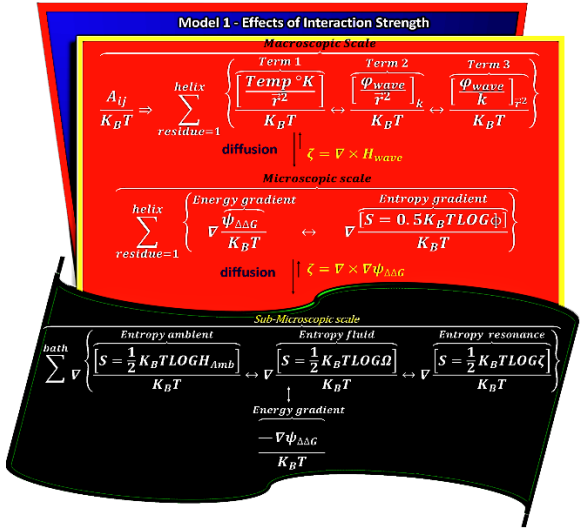

**Fig. 10. Effects of Interaction Strength.**

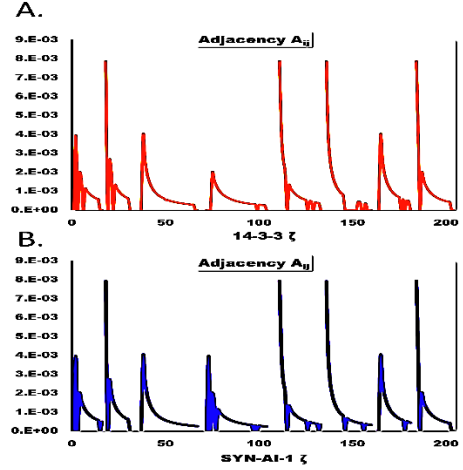

**Fig. 11 Linear Representation of Residue Adjacency $A_{ij}$.**

To consider effects of interaction strength on allostery, we solved residue adjacency $A_{ij}$ [40] as a function of background energy $\varphi$ and backbone vibrational modes in the background elastic modulus $k_\gamma$ of the helix (28). The elastic modulus $k_\gamma$ is described in respect to damping as it evolves with the decay of the Hamiltonian. The effects of energy transfer $\zeta = \nabla \times \varphi$ to lower scales does not depend on the gradient and occurs independent of damping, thusly we also accounted for its effects on interaction strength.

$$\sum A_{ij} = \frac{\varphi + \alpha - \zeta}{k_\gamma(\Delta z + \Delta l + \Delta \theta)^2} \qquad [28]$$

By incorporating residue adjacency $A_{ij}$, we account for effects of thermodynamic factors that describe the state of the oscillator in relationship to force fields $\nabla A_{ij} = \nabla Q + \nabla W$ generated by interacting residues **Fig. 10**. In its simplest form residue adjacency $A_{ij} = °kelvin/\bar{r}^2$ is an entropy factor and $|term1\rangle Q)$ describes hotspots over vibration modes created by wave propagation. Term 2 is a stability factor $|term2\rangle W\rangle$ that describes response of the vibration mode to waves $A_{ij} = [\varphi_{wave}/\bar{r}^2]_k$ in the background elastic modulus $k$ and captures the local energy distribution around vibration modes. When adjacency $A_{ij} = [\varphi_{wave}/k]_{\bar{r}^2}$ is solved according to response of the elastic modulus $k$ to standing waves, $|term3\rangle W\rangle$ and informs that allosteric motion in response to waves is dependent on the stability of the media. When thermally distributed they generate an energy signal in background energy associated with the surrounding bath. These thermodynamic factors crosstalk to reach a thermodynamic-equilibria and also transfer information across time-spatial scales as a function of their gradients. The macroscopic field drives information flow while the sub-microscopic scale acts as an information sink.

By solving adjacency as a function of background modulus $k_\gamma$ and background energy $\varphi$ with initial decays of $\alpha=0$ and

$\zeta = 0$, we found that stronger interactions occurred at helix N' terminals and connectivity decreased exponentially towards C' terminals **Fig. 11**, matching the decays of force and stiffness. Importantly, exponential decay of residue adjacency suggests genomic optimization of allostery. Based on broad exponential decay of residue adjacency in helices H3, H4 and H7, they undergo hinge-bending. Where, noise at the H4 N' terminal was due to a torsional motion. These helices were associated with entropy wells illustrated in **Fig. 9**c,d, suggesting that hinge-bending plays a significant role in allostery.

### 3.53 Allosteric Radiation and Energy Transfer
Earlier we demonstrated the relationship between free energy $\Delta G$ wells created by Z fluctuations and entropy. To consider the effects of damping and residue adjacency on the free energy landscape, we modeled allosteric free energy $\Delta\Delta G$ according to McLeish who elegantly described $\Delta\Delta G$ as the difference in two binding free energies of a ligand in the scenario the other ligand is in the bound or unbound state. Thusly, we modeled allostery based on motion in conjunction with the effects of ligand binding on the media where $\Delta\Delta G = (\beta^2/4(2\pi)^7)\bar{k}^2[k_1(x_1) + k_1(x_2)]$, $\beta^2 = 1/(K_BT)^2$ is a thermal distribution factor and allosteric free energy $\Delta\Delta G$ is considered the Hamiltonian of the resonance field. Communication between effector and allosteric residues is captured by $[k_1(x_1) + k_1(x_2)]$, whereby background elastic modulus $k$ was substituted for the expression in brackets as it describes the dimensionless moment of residues $(x_1, x_2)$ to perturbation at the effector site. Effects of residue interaction strength were incorporated by adding thermal distribution $A_{ij}/K_BT$ of residue adjacency. This allowed us to convert $\beta^2 \Rightarrow \beta^3$ and equilibrate it with $\varphi$ which was already in $R^3$. To account for effects of damping and bath interactions, we added distribution of energy decay $\alpha/K_BT$ to describe the allosteric phase shift and account for harmonic decay $\zeta$. $\Delta\Delta G$ was mapped to complex wave space (29), where each residue vibration is a separate wave function and summed over each helix.

$$\Delta\Delta G \rightarrow \sum_{AA=1}^{N} Ae^{\left(\frac{\varphi + \alpha - \zeta}{K_BT} + A_{ij}\right)}, where \; A = \frac{\pi}{8(2\pi)^7 Z_{flux}^2} \qquad [29]$$

The SCM is a coarse grain model that analyzes allostery as a function of cooperative motions of structures, thusly reduces computational stringency and noise. We modeled probability of occupation $P_o$ as illustrated in (30). Background noise was reduced by limiting residue interactions to those characterized as greatest or least in response to perturbation at the effector site with a minimum cutoff distance of 12 Å. All other sites were considered decoupled where $\varphi = 0$, $A_{ij} = 0$, $P_o = 0$ and $\alpha = 0$. For occupied states $A_{ij} = 1$, $\varphi \neq 0$, $P_o \neq 0$, $\alpha \neq 0$, where $P_o \rightarrow 1$ as $\varphi \rightarrow 0$. In all instances, the number of moles $B = 1$ and temperature $T = 273\ K$.

$$P_o = \begin{cases} 1 - \dfrac{e^{A_{ij}}}{Be^{((\varphi+\alpha-\zeta)/K_BT)+A_{ij}}} & A_{ij}=0 \quad \varphi=0 \quad \alpha=0 \\[2em] \dfrac{e^{A_{ij}}}{Be^{((\varphi+\alpha-\zeta)/K_BT)+A_{ij}}} & A_{ij}=1 \quad \varphi\neq0 \quad \alpha\neq0 \end{cases} \quad [30]$$

Communication between allosteric sites $x_1$ and $x_2$ and the effector site via allosteric waves is described in (31), where $\psi = \varphi((\varphi + \alpha - \zeta)/K_BT)$ is the thermal oscillatory energy and the second term $\emptyset = \varphi A_{ij}$ gives the allosteric phase shift due to residue adjacency. Allosteric radiation occurs by dynamically changing Hamiltonians of helices and spatial effects of residue interactions on the quantum scalar field, where $\alpha$ is dependent on the damping rate and decay $\zeta$ is continuous. Allosteric $\psi_{\Delta\Delta G}$ radiation can be described by Euler and as a wave (32).

$$\psi_{\Delta\Delta G} = \sum_{AA=1}^{N} Ae^{(\psi + \emptyset)} \quad [31]$$

$$\psi_{\Delta\Delta G} = A\sin(\psi + \emptyset)\cos(\psi + \emptyset) \quad [32]$$

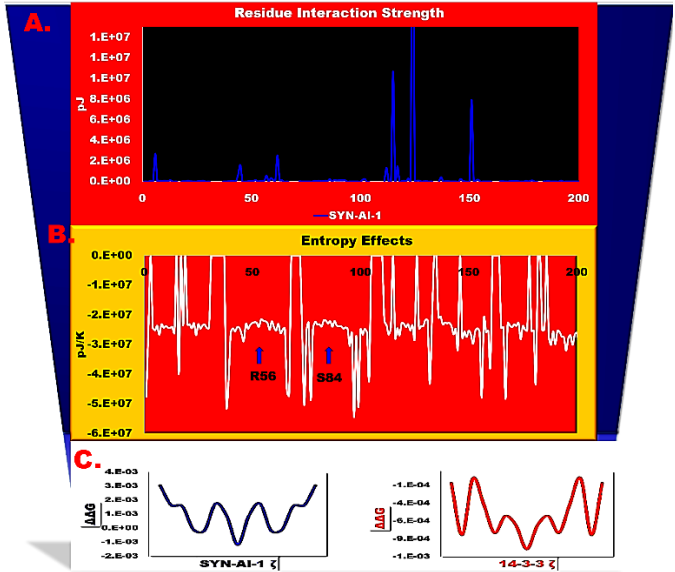

**Fig. 12 Allosteric Radiation and Free Energy**.

By describing allosteric free energy according to (31), we captured effects of residue interaction strength **Fig. 12**a on the entropy landscape **Fig. 12**b. Notably, residue adjacency plays a role in forming entropy wells, whereby when its effects were removed steepness of entropy wells was diminished **Fig. 12**b. By comparing adjacency $A_{ij}$ to entropy $S = 0.5K_BTLOG\psi_{\Delta\Delta G}$ we also noticed that interaction strength plays a role in barrier formation, especially around the well associated with effector residue R56. These relationships were verified by polynomial regressions that indicated a correlation of $R^2 = 0.41$ between $A_{ij}$ and microscopic field entropy Fig. **13**a. Interaction strength was solved by resonance entropy $S(\varphi)$ opposed to $S(\psi_{\Delta\Delta G})$, thusly was independent of $A_{ij}$ allowing us to isolate its effects. When comparing $A_{ij}$ to sub-microscopic field entropy there was a correlation of $R^2 = 30$ Fig. **13**b. These models had p-values=0 thusly establish information transfer across scales. Residual plots $y - \hat{y}$ revealed that as order increased microscopic field entropy decreased Fig. **13**c. Effects on sub-microscopic field entropy were exaggerated as entropy transfer was restricted to a small range of low strength interactions at $A_{ij} = 0 - 0.1\ Kelvin/pJ$ Fig. **13**d. Beyond that point the increased order within the media restricted entropy transfer.

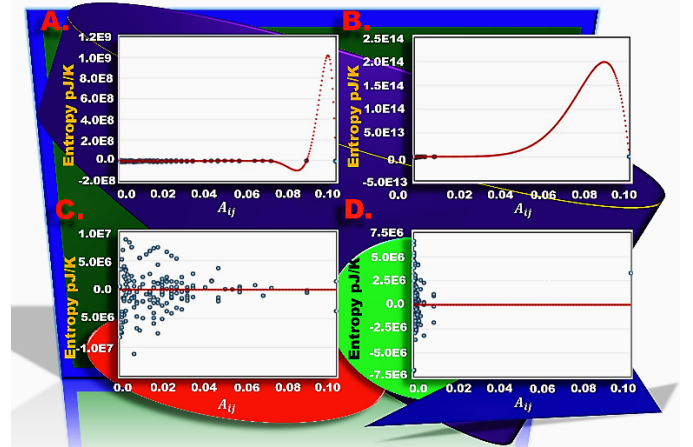

**Fig. 13. Effects of Residue Adjacency on Entropy Transfer**

Additionally, we used the allosteric free energy equation to assess effects of $\Delta\Delta G$ on ligand trapping. By summing $\Delta\Delta G$ of residues located in helices we noticed formation of $\Delta\Delta G$ wells between helix-loop-helix domains. These wells characterize the harmonic potential of docking proteins for ligand trapping. Notably, synthetic evolution changed the landscape altering the location and shape of wells in addition to barrier heights **Fig. 12**c. The Fokker-Planck equation describing probability of the $\Delta\Delta G$ scalar field is given by (33), where Langevin $L(\psi, Z) = F(\psi_{\Delta\Delta G}) + \zeta$ is a function of driving force $F(\psi_{\Delta\Delta G})$ and decay $\zeta = \nabla \times \psi_{\Delta\Delta G}$. Thusly, the ligand trapping illustrated in **Fig. 12** is dynamic and depends on field decay as well as probability of the state and the amount of driving force within the field.

$$\frac{dP(\psi, Z)}{dx} = L(\psi, Z)P_{SS}(\psi, Z) \quad [33]$$

While we did not model the sub-microscopic field in the previous section due to its incoherent drive that is a function of spatial arrangements of helices, their resonances and damping; recurrent field decay $\zeta$ of the microscopic scalar field implies a steady state flux $J_{SS}$ of free energy $\psi_{\Delta\Delta G}$ into the surrounding bath as thermal fluctuations, incompressible vortical waves and sound waves. Decay energy $\zeta$ acts as the Hamiltonian of the sub-microscopic field and its gradient $\nabla\zeta$ the driving force. The entropy $S = 0.5 K_B T Log \zeta$ landscape is a function of thermal distribution of energy transferred from the two higher time-spatial scales. Energy transfer between microscopic and sub-microscopic fields depends on diffusion coefficient tensor $D_z$. We solved $D_z$ by transforming the time-dependent classical expression of the diffusion tensor $\langle \zeta(t) \zeta(t') \rangle = 2 D_t \delta(t - t')$ to a spatial expression (36), allowing us to characterize spatial evolution of free energy flux. Effects of residue interactions on energy diffusion $D_z$ were modeled by mapping the Kronecker $\delta | A_{ij} \rangle = [0,1]$ function to adjacency space. This was done by solving $A_{ij}$ as a function of vibration mode Hamiltonians and moduli (34) opposed to background resonance $\varphi$, where for coupled residues $A_{ij} = 1$ and for decoupled residues $A_{ij} = 0$. (40) Energy flux between scales is a function of probability $P_{SS}$ of the state, the driving force and the diffusion coefficient (35).

$$A_{ij} = \frac{\varphi_z}{k_z \Delta z^2} + \frac{\varphi_\theta}{k_\theta \Delta \theta^2} + \frac{\varphi_l}{k_l \Delta l^2} \to [0,1] \qquad [34]$$

$$J_{SS} = F(\psi_{\Delta\Delta G}) P_{SS} - D_Z \frac{\partial}{\partial Z} P_{SS} \qquad [35]$$

$$\langle \zeta(Z) \zeta(Z') \rangle = 2 D_Z A_{ij}(Z - Z') \qquad [36]$$

### 3.5.4 Modelling Allostery Along an Ideal Stretched String

To show conversion of the scalar field to allosteric pressure waves flowing through the elastic media, we modeled Langevin $\nabla_{\Delta \vec{r}, t} \psi_{\Delta\Delta G} = F(\psi_{\Delta G}) + \zeta$ along an ideal stretched string where decay of allosteric free energy to entropy was captured by $\zeta$. Driving force $F(\psi_{\Delta\Delta G})$ was a function of tension $\tau_S$ along the ideal stretched string and the spatial evolution of the elastic media. Due to small $\alpha$ angles created by the tension vector [41] and the convolution of the media we solved velocity along the string as the tangent $\partial Z / \partial \Delta \vec{r}$ of $Z$ fluctuations in respect to backbone vibration modes. In the time dimension, $F(\psi_{\Delta\Delta G})$ was a function of mass $m = \mu \Delta l$ and acceleration of the allosteric wave along the string, where $u$ is the density of the elastic media and $l$ structure length. We modeled the protein surface as an isotropic and homogeneous membrane, thusly in the space dimension motion simplified to tangential displacement of allosteric waves on the stretched string, where $\partial Z_1$ and $\partial Z_2$ describe velocity at the crest and base of waves (37).

$$\overbrace{\tau_S \left( \frac{\partial Z_1}{\partial \Delta \vec{r}} - \frac{\partial Z_2}{\partial \Delta \vec{r}} \right)}^{F_{\Delta \vec{r}}(\psi_{\Delta\Delta G})} = \overbrace{\mu \Delta l \left( \frac{\partial^2 Z}{\partial t^2} \right)}^{F_t(\psi_{\Delta\Delta G})}$$

$$\Rightarrow \overbrace{\left( \frac{\partial Z_1}{\partial \Delta \vec{r}} - \frac{\partial Z_2}{\partial \Delta \vec{r}} \right)}^{space} = \overbrace{\left( \frac{\partial^2 Z}{\partial t^2} \right)}^{time} \qquad (37)$$

By taking the limit as waves approach the minimum energy conformation as backbone displacement $\Delta \vec{r} \to 0$, we show that transduction of allosteric information is solely a function of wave acceleration along the ideal stretched string (38). $\Delta \vec{r}$ is dimensionless thusly waves radiate in all directions where configuration of the entropy landscape and flow of allosteric motion depend on docking protein three-dimensional structure.

$$lim_{\Delta Z \to 0} \Rightarrow \frac{1}{\Delta \vec{r}} \left( \frac{\partial Z_1}{\partial \Delta \vec{r}} - \frac{\partial Z_2}{\partial \Delta \vec{r}} \right) = \left( \frac{\partial^2 Z}{\partial \Delta \vec{r}^2} \right)$$

$$\Rightarrow \overbrace{\left( \frac{\partial^2 Z}{\partial \Delta \vec{r}^2} \right)}^{F_x(\psi)} = \overbrace{\left( \frac{\partial^2 Z}{\partial t^2} \right)}^{F_t(\psi)} \qquad (38)$$

By splitting the above operators, allosteric driving force $F(\psi_{\Delta\Delta G}) \Rightarrow f_1(\Delta \vec{r} - t) + f_1(\Delta \vec{r} + t)$ created by the free energy gradient $\nabla_{\Delta \vec{r}, t} \psi_{\Delta\Delta G}$ is implicitly a function of space and time dimensions of standing waves passing through the continuous elastic media. Where, function $f_1 = k$ describes the effect of the elastic modulus on wave transduction.

$$\left( \frac{\partial Z}{\partial \Delta \vec{r}} - \frac{\partial Z}{\partial t} \right) \left( \frac{\partial Z}{\partial \Delta \vec{r}} + \frac{\partial Z}{\partial t} \right) \Rightarrow f_1(\Delta \vec{r} - t) + f_1(\Delta \vec{r} + t) \qquad (39)$$

Wave functions $A cos(\Delta \vec{r} - t)$ and $A cos(\Delta \vec{r} + t)$ describe motion in opposing directions, where amplitude $A$ is the Z fluctuation of the elastic media in response to perturbation. The QHO gradient is $\nabla_{\Delta \vec{r}, t} \psi \cong k \Delta \vec{r} + m \omega$, where driving force $k \Delta \vec{r}$ characterizes the spatial component and the time component simplifies to momentum $m \omega$ (40). Whereby, the allosteric phase shift is $\alpha = \nabla_{\Delta \vec{r}, t} \emptyset$. [3] Pressure waves emanating from oscillating helices at the microscopic scale are a function of allosteric free energy gradient $\nabla_{\Delta \vec{r}, t} \psi_{\Delta\Delta G}$ and the allosteric phase shift with respect to surface area (41).

$$\nabla_{\Delta \vec{r}, t} \psi_{\Delta\Delta G} = A sin(k \Delta \vec{r} + m \omega + \alpha) cos(k \Delta \vec{r} + m \omega + \alpha) \qquad (40)$$

$$P_{\Delta\Delta G} = P_A sin(\nabla_{\Delta \vec{r}, t}(\psi + \emptyset)) cos(\nabla_{\Delta \vec{r}, t}(\psi + \emptyset)) \qquad (41)$$

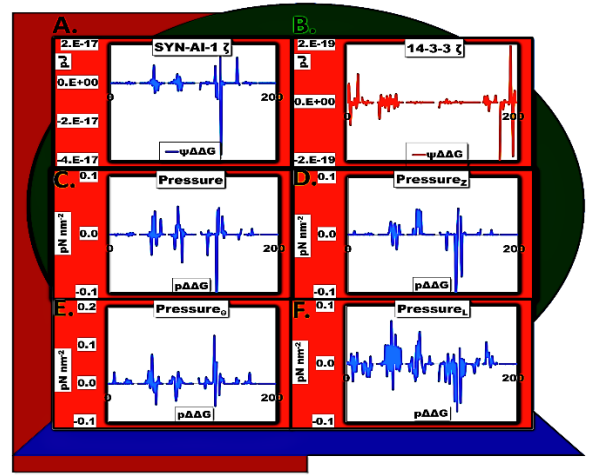

**Fig. 14 Pressure Waves along Ideal Stretched String**

Based on data from **Fig. 14**c-f, it is obvious that allosteric pressure waves $P_{\Delta\Delta G}$ due to backbone vibration modes are a function of ground state Hamiltonian $\psi_{\Delta\Delta G}$ **Fig. 14**a. We confirmed coil-like behaviors of amphiphilic residue blocks in helices H3 and H4 as well as residues V132 and D137 of H5 and H6 based on spikes in their scalar fields **Fig. 14**a,b. This was supported by concurrent decreases in pressure due to the associated loss of driving force in these regions **Fig. 14**c,f.

Microscopic entropy changes in elastic media surrounding helices are a function of pressure waves $P_{\Delta\Delta G}$ emanating from helical oscillations. $P_{\Delta\Delta G}$ waves effect the entropy landscape by the relationship $S = 0.5K_BTLogP_{\Delta\Delta G}$. During decay pressure waves convert to thermal fluctuations with some energy lost as sound waves $H_{sound} = \mu\Delta\hat{\pi}c^2$, where $\mu$ describes the local density of the continuous elastic media and velocity $c^2 = \left[\partial P_{\Delta\Delta G(3)}/\partial Z\right]_S$ is a function of entropy $S$ along the stretched molecular string. We corroborated our earlier finding as entropy increases near residues R56, S84, V132 and D137 **Fig. 15**c led to decrease of allosteric free energy $\Delta\Delta G$ **Fig. 15**a and promoted routing of allosteric pathways through helices H3 and H4. Residues V132 and D137 are located in an entropy well at the H5-C5-H6 helix-to-coil transition **Fig. 1**f. Notably, changes in $\Delta\Delta G$ correlated with those in elastic modulus $k$ illustrated in **Fig. 9**a,b. The drop in $\Delta\Delta G$ explains the loss of driving force in these regions.

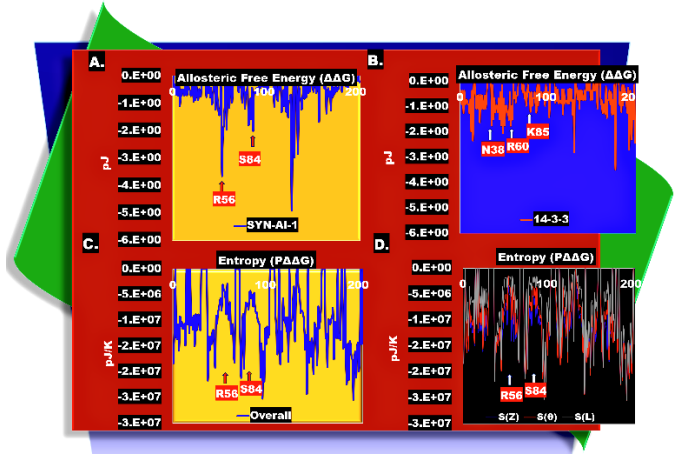

**Fig. 15 Allosteric Free Energy and Entropy Landscapes at the Microscopic scale.** Free energy landscape of SYN-AI-1 $\zeta$ (**A**) and 14-3-3 $\zeta$ (**B**). Entropy landscape of SYN-AI-1 $\zeta$ (**C**). Entropy due to pressure waves emanating from tangential, rotational and stretching modes (**D**).

At the sub-microscopic scale quantum theory was not applied due to white noise generated by its incoherent drive. The drive has three primary fueling sources, $\Omega$ donated by the helical power stroke, $\phi$ of helices, and thermal energy from the ambient environment of the cell, thusly Langevin $\nabla H_{sMi} \Rightarrow \nabla_\Omega + \nabla\phi + \nabla_{Amb}$ is a composite of three interacting gradients. Transferred $\Omega$ is in constant decay at the end the conformational switch, thusly resonance $\phi$ and ambient energy are the primary driving force. Ambient energy is continuous and functions as the background energy, whereby resonance $\phi$ undergoes decay due to damping. Allostery may be experimentally characterized by hardly discernable local temperature increases and decays in the background energy at small time intervals, where pressure $P_\zeta \sim \sum_{AA=1}^N P_A sin(\psi + \emptyset) cos(\psi + \emptyset)$ waves emitted from the oscillating string enter the sub-microscopic field and generate heat spikes as a function of their thermal distribution $\psi = P_\zeta\left(P_\zeta/K_BT\right)$, where $P_\zeta$ is described by (42) and $\zeta$ by (43). Allosteric phase shift $\emptyset \Rightarrow \nabla\Im = \left[\nabla(\nabla \times H_{sMi}\right]_{A_0}$ describes the decay of pressure due to heat dissipation $\Im = \nabla \times H_{sMi}$ and unit vectors $\hat{\imath}$, $\hat{\jmath}$, and $\hat{k}$ describe the direction of vibration modes. Energy dynamics vary at this scale as the flow of information is not spatially confined to helix resonances but described by a thermal bath. Pressure waves enter the field and disperse until thermodynamic equilibrium is obtained creating local entropy gradients. Crosstalk with the microscopic field is shown in the form of field entropy $S = \left[0.5K_BTLOGP_\zeta\right]_\varphi$ with respect to helix resonance **Fig. 16**. Whereby, we observed formation of entropy wells in the proximity of effector residues E28, R56, S84, V132 and D137. The shape and depth of entropy wells and barriers varied from the microscopic field.

$$P_\zeta = [\nabla(\nabla \times \psi_{\Delta\Delta G})]_{A_0} \Rightarrow \boldsymbol{\nabla}\begin{vmatrix} \dfrac{\widehat{\tau}}{d\Delta z} & \dfrac{\widehat{J}}{d\Delta\theta} & \dfrac{\widehat{k}}{d\Delta l} \\ |d\Delta z| & |d\Delta\theta| & |d\Delta l| \\ \dfrac{\partial}{\partial\Delta z} & \dfrac{\partial}{\partial\Delta\theta} & \dfrac{\partial}{\partial\Delta l} \\ \psi_{\Delta z} & \psi_{\Delta\theta} & \psi_{\Delta l} \end{vmatrix}_{A_0} \quad (42)$$

$$\zeta$$
$$\left(\frac{\partial\Psi_l}{\partial\Delta\theta} - \frac{\partial\Psi_\theta}{\partial\Delta l}\right)\hat{\imath} + \left(\frac{\partial\Psi_z}{\partial\Delta l} - \frac{\partial\Psi_l}{\partial\Delta\theta}\right)\hat{\jmath} + \left(\frac{\partial\Psi_\theta}{\partial\Delta z} - \frac{\partial\Psi_z}{\partial\Delta\theta}\right)\hat{k} \quad (43)$$

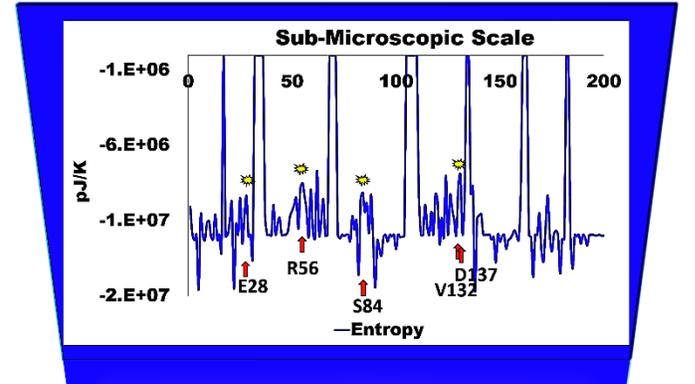

**Fig. 16 Entropy Landscape at the Sub-Microscopic scale.** A formula for curl $y = [\nabla \times \psi_{\Delta\Delta G}]_\varphi$ was solved for as a function of helix $\varphi$ using a 11[th] order polynomial regression where calculated curl (42) matched predicted curl at $R^2 = 0.99$. The formula was used to solve $F_\zeta = \nabla[\nabla \times \psi_{\Delta\Delta G}]_\varphi$. Entropy $S = 0.5K_BTLOGP_\zeta$ was calculated as function of sub-microscopic pressure waves where force $F_\zeta$ was considered in respect to residue surface area.

## Discussion

In the current study, we utilized the Spatial Convolution Model (SCM) to investigate the effects of synthetic evolution [11, 12] on allostery within 14-3-3 $\zeta$ docking proteins. SCM approximates motion based on wave dimensionless moments (Z fluctuations) solved as a function of greatest and least residue motions in response to a perturbation at a distant effector site. Convolution of the continuous elastic media due to allosteric motions was solved as a function of Z fluctuations in respect to backbone vibration modes. Allosteric motions were analyzed using the elastic network model (ENM), a coarse grain method that does not account for thermal factors but is very accurate in approximating force gradients. [27, 28] We modeled allostery as a three-step process involving I) a macroscopic step wherein allostery occurs by standing waves produced by helix power strokes during conformational switching, II) a microscopic step characterized by structural oscillations whose decay transfers energy to III) a microscopic step characterizing driving force of thermal fluctuations associated with motion free allostery.

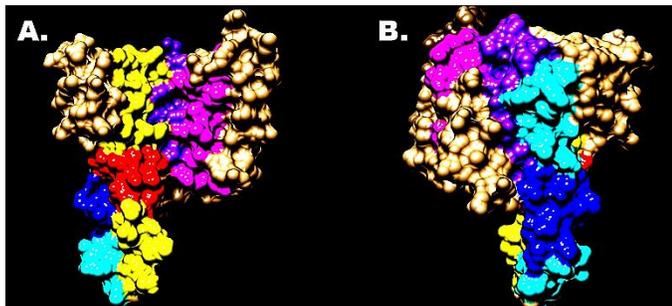

**Fig. 17. Spatial Association of Hinge-bending and Active Site in SYN-AI-1 ζ.** Conserved residues described in [12] located in the amphipathic groove are colored purple and magenta, helix H3 is yellow and disordered region residues 52-60 are red (**A**). Helix H4 is located dorsally to H3 and colored light blue and disordered region residues 80-91 are dark blue.

We found that transfer of allosteric information in 14-3-3 ζ docking proteins occurs by standing waves **Fig. 1**a,b and were able to visualize entropy gradients as a function of standing wave Hamiltonians. Entropy gradients and wells formed as a function of the orientation and combined motions of helices and were characterized by a definite slope indicating the direction of the flow of information **Fig. 1**e,f. However, due to synthetic evolution the trend was altered in the synthetic docking protein as entropy wells formed independently in helices H3 and H4 due to the context of amphiphilic residues. Reversal in standing wave direction led to a reversal of the entropy gradient **Fig. 1**e,f and the flow of allosteric information. Based on localization of coils in entropy wells and increases in entropy $S \to 0$, they act as allosteric hotspots due to low conformational stiffness. In the native protein intradomain coils were entropically stable, while interdomain coils connecting helix-loop-helix domains localized in entropy wells. This suggests that interdomain coils play a role in entropic changes that contribute to formation of potentials that contribute to ligand binding, and that during the rotation of structural domains when these potentials are created intradomain coils are responsible for domain stability. This would imply that binding pockets are formed between structural domains. Additionally, allostery was sequence dependent with amphiphilic residues located at effector and allosteric sites which resulted in a local reduction of stiffness and routing of allostery through helices.

Notably, the evolution strategy used by synthetic evolution was similar to retroviral proteases in achieving drug resistance [42], where mutations in helices displaying hinge-bending alter but do not disrupt active site geometry. Mutations in helix H3 associated with the disordered region from residues 52-60 were introduced below and adjacent to the amphipathic groove **Fig. 17**a, thusly they affect its entropy landscape during hinge-bending motions but do not distort its shape. This was also true of mutations in helix H4 **Fig. 17**b, as amphiphilic residues 80-91 associated with structural disorder and hinge-bending were located behind, below and adjacent to the amphipathic groove. Importantly, the increased plasticity in these regions increases the number of conformational states of the groove broadening the potential ligand binding pool. Our findings not only suggest such sites are excellent targets for altering allostery but also for drug design.

Although, the continuous elastic media of docking proteins is a complex three-dimensional structure, allostery behaved as if it were conducted along a stretched oscillating string whereby local entropies generated by evolutionarily optimized energy gradients affect allosteric wave velocity. This was corroborated by increased wave velocities through helices H3 and H4 **Fig. 5**f of the synthetic homodimer. Behavior of the elastic media as an oscillating string was reflected by resonance patterns of wave Hamiltonians that were damped harmonic oscillators **Fig. 1**c,d. Inhomogeneities along the oscillating string affected allostery as Hamiltonian $\Delta E$ of the helix power stroke matched C-alpha strain across the docking protein **Fig. 2**, suggesting that helix mechanical dynamics shape the energy landscape. Resonance energy $\phi$ resulting from fast decomposition of the macroscopic scalar field also matched the pattern of C-alpha strain **Fig. 6**. Additionally, the behavior was supported by effects of the loss of driving force in helices H3 and H4 on the oscillation of the string due to a loss of local tension **Fig. 3**a, which decreased pressure waves emanating from the string **Fig.14**c-f. Notably, the string overlapped at structural domains in an evolutionarily optimized manner that allowed optimal transfer of energy.

Importantly, allostery within macroscopic and microscopic scalar fields depends on structural inhomogeneities within the oscillating string. This became apparent with the introduction of amphiphilic sequence blocks near effector residues R56 and S84, which induced coil-like behavior allowing allostery to be routed through helices H3 and H4 in SYN-AI-1 ζ **Fig. 1**f and **Fig. 9**c. Allostery was promoted by introduction of entropy wells along the string by amphiphilic residues that resulted in symmetry breaking. [43, 44] The entropy of both the fields were similarly affected by mutation. Entropy increases at effector sites correlated with decreased driving force $F(\Psi_H)$ supplied by the macroscopic scalar field gradient $\nabla H_{Ma}$ **Fig. 3**a. While synthetic evolution disrupted force gradients, harmonic tuning of energy gradients **Fig. 4**a and damping **Fig. 4**a were not disrupted only modified suggesting they are very robust. Symmetry breaking in helix H3 of SYN-AI-1 ζ occurred near residues R56 and S57. Serine acts as a helix disruptor that allows fast propagation of allosteric information by bending α-helices in the $\chi_1 = g^1$ conformation resulting in a decrease in the ∅ torsion angle and an increase in the ψ torsion angle. [45] Symmetry breaking in helix H4 occurred at residue S84 in the context of a chain of amphiphilic residues (R80, R81, D83, K85, E87, E89, R91).

In the native 14-3-3 ζ homodimer residues N38, R60, and K85 routed allosteric pathways through helices H3 and H4 **Fig. 9**d. While effector residues localized to disordered regions, allosteric sites were in both ordered and disordered regions, but in all cases were associated with low stiffness supporting that allostery depends on the potential energy landscape. Notably, native entropy wells formed without disruption of driving force suggesting evolutionary optimization **Fig. 3**b. Wells were small compared to those in synthetic helices and undetectable at the macroscopic scale which corroborates evolutionary optimization of entropy well formation. As in the synthetic docking protein they encompassed effector sites and were associated with amphiphilic sequences. The entropy well associated with

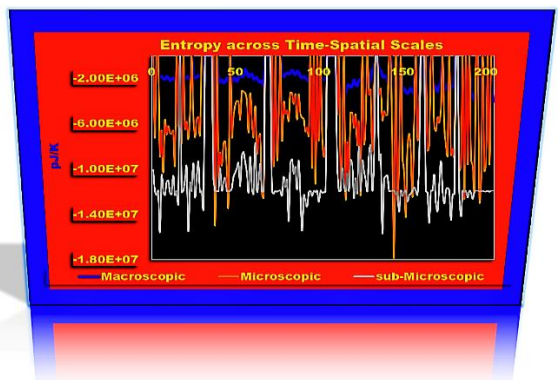

**Fig. 18 Entropy across Time-Spatial Scales**

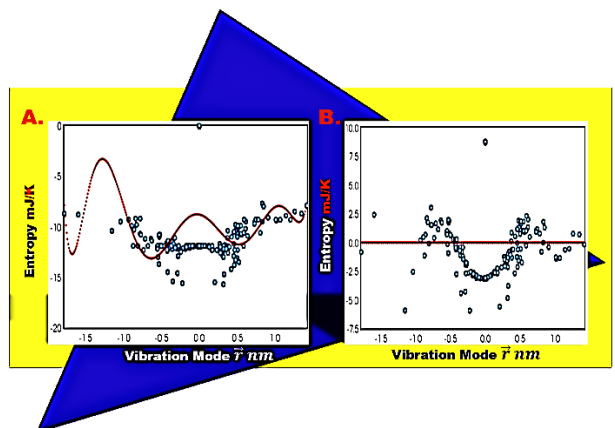

**Fig19. Formation of the Ligand Binding Pocket.** Residue vibration mode $\vec{r}$ was compared to sub-microscopic field entropy by polynomial regression using the Blue.stats online regression tool (**A**) which was also used to generate the $y - \hat{y}$ residual plot (**B**). The model was verified with a correlation score of $R^2 = 0.89$ and a p-value=0.

effector residue N38 was in the context of (S37, N38, E39, E40, R41, N42), the well near R60 was in the context of (Y59, R60, V61), and the well near residue K85 was in context (D83, S84, I86, E87, T88, E89). The presence of serine in these sequence blocks suggests nature uses the residue to reduce local stiffness and route allostery through helices.

As mentioned earlier, we propose that allostery occurs at three major scales that each comprise of an infinite number of quantum states at varying probabilities. Impacts of mutation vary across these scales as characterized by entropy landscapes illustrated in **Fig. 18**. At the macroscopic scale entropy is a function of gradient $\nabla H_{Ma}$ produced during conformational switching and described by the Langevin $\nabla_z H_{Ma} = F(\Psi_H) + f(\phi + \Omega)$, where $\Psi_H$ is the internal energy generated during the motion. Importantly, combined motions of helices at this scale created free energy $\Delta G$ wells between Z fluctuation amplitudes that generated a rolling entropy landscape as standing waves travelled the docking protein. At the microscopic scale entropy is a function of macroscopic field decay Hamiltonians ($\phi, \Omega$) that act as its driving force $f(\phi + \Omega)$. Energy transfer at this scale occurs in a continuous collision free elastic media and can range from molecular to sub-electron, [46] thusly we quantized the field by $|\phi\rangle\psi_{\Delta\Delta G}\rangle$ mapping resonance $\phi$ to the quantum field Hamiltonian $\psi_{\Delta\Delta G}$. Allosteric free energy $\Delta\Delta G$ was solved according to McLeish which allowed us to consider damping and interactions with the surrounding bath as well as effects of residue interaction strength. [26] By doing so, we found that residue interaction strength effected shape and depth of entropy wells as well as barrier heights. Spatial effects of higher scales are invisible to the sub-microscopic field which is comprised of a continuous bath of energy from the surrounding environment. Energy $\zeta$ transferred as pressures waves from resonating helices enter the field and create driving force of thermal fluctuations that create short-lived hotspots in the background energy and local entropy gradients **Fig. 18** that should be experimentally detectable. Notably, loss of driving force at the macroscopic scale resulted in free energy decreases **Fig. 15**a and formation of entropy wells **Fig. 15**c in the oscillating string that decreased pressure waves **Fig. 14**c-f and resulted in a loss of energy flux into the sub-microscopic field.

Allostery at long distance requires inhomogeneities in the continuous elastic media that allow formation of helical traps, where fluctuations in a surrounding thermal bath form potential energy wells trapping ligands in a harmonic potential implied by a Hookean spring constant $k$ representing stiffness of the trap. [9, 47] Based on our findings, $\Delta\Delta G$ wells form between neighboring helix-loop-helix domains by the combined motion of domains around interdomain coils as helices forming these domains have similar elastic moduli $k$. However, neighboring domains comprise of nonuniform elastic moduli **Fig. 9**a,b that exists in a surrounding oscillatory $\psi_{\Delta\Delta G}$ bath comprised of large gradients **Fig. 1**c,d between structural domains that allow for helical trapping. Importantly, we were able to provide evidence for formation of binding wells by cooperative motion and long-range helix interactions by comparing backbone vibration mode $\vec{r}$ created during standing wave propagation to sub-microscopic field entropy **Fig. 19**a, whereby the submicroscopic field acts as an entropy sink. We discovered a binding pocket was formed by the organization of high entropy residues. Notably, the trend line shows wave propagation across the docking protein and provides clear evidence that entropy well formation is created by wave standing waves. We uncovered the mechanism of well formation by residual plotting which allowed visualization of well depth and breadth **Fig. 19**b. A small binding pocket also formed to the right of the well that could possibly allow for the binding of a cofactor. Branches extending from the bottom of the binding pocket were associated with low entropy residues that likely characterize regions of the docking protein that are responsible for structural support during ligand binding.

Ligand binding pocket formation is based on the structural organization of high entropy residues thusly depends on elastic modulus $k$ which is a function of tangential, rotational, and stretching mode Hamiltonians that dynamically change due to helix damping. Therefore, the entropy landscape shown in **Fig. 19** has both a spatial and time evolution. We demonstrated earlier that helices have different damping coefficients **Fig. 4**, thusly stiffness of helical traps and shape of the binding pocket change as the media evolves. The surrounding oscillatory $\psi_{\Delta\Delta G}$ bath evolves as a function of damping and a continuous energy exchange with the surrounding bath. Thusly, conditions of the helical trap are altered where it is more advantageous for some ligands to interact at specific time-spatial intervals. Importantly, we established that information transfer between scalar fields plays a role in binding pocket formation. Information transfer implies that structural resonance ϕ, fluid Ω driving forces, field decay rates and driving force are dynamic. Thusly, energy flux $J_{SS}$ is not a steady state and comprises of a constantly changing diffusion constant $D_Z$.

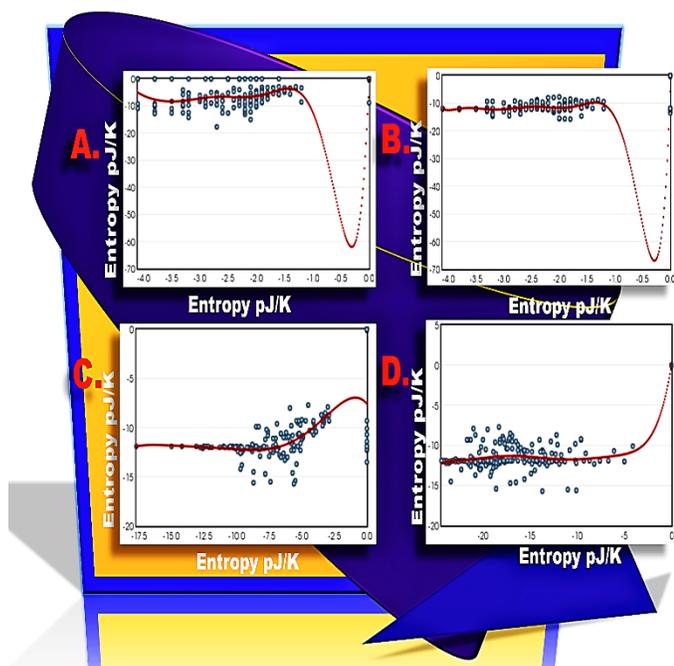

**Fig.20 Entropy Transfer across Scales.** To investigate information transfer across time-spatial scales polynomial regressions were performed using the Blue.Stats online regression tool. We compared macroscopic field entropy to microscopic field entropy (**A**), macroscopic field entropy to sub-microscopic field entropy (**B**), microscopic field entropy to sub-microscopic field entropy (**C**), rolling entropy to sub-microscopic field entropy (**D**). All models were significant with p-values=0.

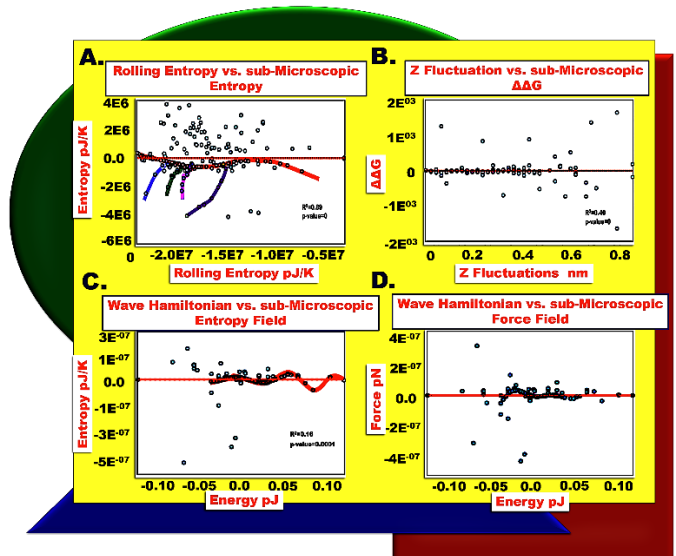

**Fig.21 Entropy diffusion Mechanisms.** To investigate information transfer mechanisms residual plots $y - \hat{y}$ were performed using the Stats.Blue online polynomial regression tool. All models were significant with p-values=0.

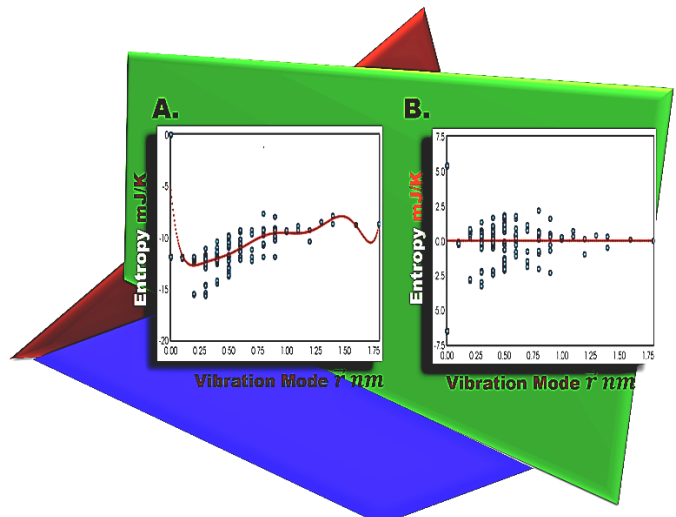

**Fig. 22 Discrete Information Transfer**

As expected of a diffusion process, entropy transfer was multi-directional and heteroscedastic as verified by polynomial regression and F-tests. Entropy transfer from the macroscopic to microscopic field had a strong correlation of $R^2 = 0.46$ **Fig. 20**a and entropy transfer from the macroscopic field to the sub-microscopic was demonstrated at an $R^2 = 0.39$ **Fig. 20**b. A correlation of $R^2 = 0.26$ describing entropy transfer from the microscopic field to the sub-microscopic scale eliminated it as the driver and suggests rolling entropies created by standing wave propagation are the primary entropy driver. However, the model displayed well fitted data and a p-value=0 confirming entropy transfer between the fields **Fig. 20**c. We corroborated the macroscopic field as the entropy driver by comparing rolling entropy to sub-microscopic field entropy **Fig. 20**d. The relationship had an identical confidence as the correlation between the backbone vibration mode $\vec{r}$ and sub-microscopic entropy with an $R^2 = 0.89$ and a p-value of 0. Thusly, we confirmed that rolling entropies are the primary entropy fueling source. Additionally, we validated multi-directional flow of information by establishing entropy transfer from the sub-microscopic field to the macroscopic at $R^2 = 0.34$ and from the sub-microscopic to the microscopic scalar field at an $R^2 = 0.23$. Our models were significant and non-random as they were characterized by p-value=0.

Entropy transfer between time-spatial scales occurred by multiple processes, a predominant diffusion process associated with standing wave propagation along the protein backbone, and peripheral processes characterized by branch points that capture diffusion processes of helices **Fig. 21**a. Notably, we visualized white noise due to overlapping diffusion processes. Noise was only located in the positive region, suggesting that diffusion is dependent on the spatial orientation of the docking protein. Effects of the standing wave Hamiltonian on entropy diffusion were illustrated in **Fig. 21**c. There was a strong trend along the protein backbone capturing propagation of the wave as well as a diffusion process at lower energy levels. Likewise, there was an increase in diffusion as driving force decreased **Fig. 21**d. The state of the standing wave as it propagates the docking protein is connected to internal energy of the docking protein by the First Law of thermodynamics $\nabla H_{wave} = \nabla Q + \nabla W$. Low diffusions at high Hamiltonians were due to a shift toward work which limited the amount of entropy available for diffusion. We observed multiple dynamics as diffusion did not obey the amount of order in the media and increased at shorter vibration modes **Fig. 22**. It followed the entropy gradient and was dependent on the standing wave Hamiltonian. Increased diffusion at shorter vibration modes was due to association of the elastic media with less energetic regions of the wave as indicated by the lower slope and by the pattern of diffusion observed in **Fig. 21**c. This was supported by increased entropy diffusion with stretching of the elastic media as indicated by Z fluctuations **Fig. 21**b, where combined effects of stretching motions obeyed the elastic modulus. Thusly, while diffusion at the microscopic scale was dominated by background elastic modulus $k$ at the macroscopic scale it was a combined function of the response of the elastic modulus to the Hamiltonian of the standing wave.

Interestingly, the complex multi-dimensional elastic media of the docking protein behaved like an oscillating string, thusly we theoretically proofed SCM by modeling the allosteric free energy gradient $\nabla_{\Delta \vec{r}, t} \psi_{\Delta \Delta G} = f_1(x - t) + f_1(x + t)$ along the ideal stretched string described in Elmore [41]. When we simplified its motion equation energy transfer was a function of space and time dimensions of allosteric waves traversing the string, with $f_1$ denoting background elastic modulus $k$ of the helix which suggests that inhomogeneities in the string affect allosteric waves. By taking the limit of the motion equation as waves approach the minimum energy conformation $\Delta \vec{r} \to 0$, we proved that allostery was a function of wave acceleration along the ideal stretched string. Additionally, by expressing allostery as a function of the free energy gradient $\nabla_{\Delta \vec{r}, t} \psi_{\Delta \Delta G}$ along the string we demonstrated information transfer by pressure waves and established that local changes in the gradient affect pressure waves emanating from the oscillating string. These oscillations supply the driving force for allostery and shape the entropy landscape. Notably, we found that pressure waves radiate from the string and diffuse across scalar fields by a diffusion gradient dictated by the standing wave Hamiltonian. Notably, allostery can be re-engineered by processes such as synthetic evolution [11, 12] that introduce inhomogeneities within the oscillating string and alter allostery by reducing local tension.

## CONCLUSION

In the current study, we used the Spatial Convolution Model (SCM) to analyze allostery in 14-3-3 ζ docking proteins. The model is advantageous as input data is acquired from ENM force fields that are very accurate, thusly SCM can approximate allostery without a need for direct experimental observation. By applying the SCM, we demonstrate that allosteric information traverses 14-3-3 ζ docking proteins as standing waves that can be characterized by a Fourier series and that synthetic evolution rewired allostery by altering standing wave dynamics. Allostery was rewired in a sequence specific manner, where localization of amphiphilic residues promoted routing of allosteric pathways through helices in native and synthetic docking proteins. We also established that allostery behaves as an energy transfer process occurring across a stretched oscillating molecular string that overlaps at domain formations. Whereby, nonuniformities in structural mechanical properties introduce inhomogeneities that drive allostery by creating local entropy gradients along the string and free energy gradients responsible for helical trapping.

**Acknowledgments** Proteins analyzed in this study were designed using the Stampede 2 supercomputer supplied by the Texas Advanced Computing Center (TACC) at the University of Texas at Austin who provided HPC resources.

**Funding** This project was funded by private contributors.

## Compliance with ethical standards

**Conflict of interest** The author declares he has no conflict of interest.

**Ethical approval** The article does not contain any studies with human participants or animals performed by the author.